\documentclass[fleqn,usenatbib]{mnras}

\usepackage{newtxtext,newtxmath}

\usepackage[T1]{fontenc}

\DeclareRobustCommand{\VAN}[3]{#2}
\let\VANthebibliography\thebibliography
\def\thebibliography{\DeclareRobustCommand{\VAN}[3]{##3}\VANthebibliography}

\usepackage{graphicx}	
\usepackage{amsmath}	
\usepackage{lscape}
\usepackage{color}
\usepackage{longtable}
\usepackage{supertabular}
\usepackage[dvipsnames]{xcolor}

\newcommand{\ee}[1]{\mbox{${} \times 10^{#1}$}}
\newcommand{\lbol}{\mbox{$L_{\rm bol}$}} 
\newcommand{\lint}{\mbox{$L_{\rm int}$}} 
\newcommand{\lintfit}{\mbox{$L_{\rm int}^{\rm fit}$}} 
\newcommand{\lintfitc}{\mbox{$L_{\rm int}^{\rm fit,corrected}$}} 
\newcommand{\lintmodel}{\mbox{$L_{\rm int}^{\rm model}$}} 
\newcommand{\lacc}{\mbox{$L_{\rm acc}$}} 
\newcommand{\lext}{\mbox{$L_{\rm ext}$}} 
\newcommand{\lstar}{\mbox{$L_{\rm star}$}} 

\newcommand{\degree}{\mbox{$^{\circ}$}}

\newcommand{\um}{$\mu$m}
\newcommand{\lsun}{\mbox{L$_\odot$}}
\newcommand{\msun}{\mbox{M$_\odot$}}

\newcommand{\mcore}{\mbox{$M_{\rm core}$}}
\newcommand{\mdisc}{\mbox{$M_{\rm disc}$}}
\newcommand{\mstar}{\mbox{$M_{\rm star}$}}
\newcommand{\rcore}{\mbox{$r_{\rm core}$}}
\newcommand{\rdisc}{\mbox{$R_{\rm disc}$}}
\newcommand{\mdotstar}{\mbox{$\dot{M}_{\rm star}$}}
\newcommand{\mdotdisc}{\mbox{$\dot{M}_d$}}
\newcommand{\fseventy}{\mbox{$F_{70}$}}

\newcommand{\fhundred}{\mbox{$F_{100}$}}
\newcommand{\fhsixty}{\mbox{$F_{160}$}}
\newcommand{\sdl}{\mbox{$\sigma_{\delta}$}}
\newcommand{\sdlu}{\mbox{$\sigma_{\delta}^{\rm updated}$}}
\newcommand{\fnu}{\mbox{$F_{\nu}$}}

\title[Estimating protostellar luminosities]{Estimating the Luminosities of Protostars with Limited Infrared Photometry}

\author[M. M. Dunham et al.]{
Michael M.~Dunham$^{1}$\thanks{E-mail: mdunham@middlebury.edu}, 
Aina Palau$^{2}$, 
Nuria Hu{\'e}lamo$^{3}$, 
Eduard I. Vorobyov$^{4,5}$, 
Zach Yek$^{6}$, \&
Sean Rand$^{6}$
\\
$^{1}$ Department of Physics, Middlebury College, Middlebury, VT 05753, USA \\
$^{2}$ Universidad Nacional Aut\'onoma de M\'exico, Instituto de Radioastronom\'ia y Astrof\'isica, Antigua Carretera a P\'atzcuaro 8701, Ex-Hda. San Jos\'e de la Huerta, \\58089, Morelia, Michoac\'an, M\'exico \\
$^{3}$ Centro de Astrobiolog\'{\i}a (CAB), CSIC-INTA,  ESAC Campus, Camino bajo del Castillo s/n, E-28692 Villanueva de la Ca\~nada, Madrid, Spain \\
$^{4}$ Institut f\"ur Astro- und Teilchenphysik, Universit\"at Innsbruck,
Technikerstra{\ss}e 25, 6020 Innsbruck, Austria \\
$^{5}$ University of Vienna, Department of Astrophysics, T\"urkenschanzstrasse 17, 1180, Vienna, Austria \\
$^{6}$ Department of Physics, State University of New York at Fredonia, 280 Central Avenue, Fredonia, NY, 14063, USA
}

\date{Accepted XXX. Received YYY; in original form ZZZ}

\pubyear{\the\year{}}

\begin{document}
\label{firstpage}
\pagerange{\pageref{firstpage}--\pageref{lastpage}}
\maketitle

\begin{abstract}
The luminosities of protostars provide one of the only indirect methods of measuring their masses and mass accretion rates in their earliest stages of evolution. Accurate measurements of protostellar luminosities traditionally requires assembling complete spectral energy distributions (SEDs) from the near-infrared through millimeter wavelengths.  In this work, we use published evolutionary radiative transfer models of collapsing protostellar cores to evaluate the extent to which protostellar luminosities can be estimated from a limited number of infrared photometric measurements.  We confirm previous results showing a tight correlation (in log-log space) between the luminosity of a protostar and its flux at 70 microns, although we demonstrate that these previous results yield luminosity estimates that are too low by factors of 2--3. We expand this work to additional wavelengths, finding that single wavelengths at 40~\um~--~350~\um\ provide luminosity estimates with a 1$\sigma$ uncertainty of a factor of 3 (0.477 dex of \lsun) or lower, with the uncertainty reduced to a factor of 2 (0.301 dex of \lsun) or lower at 70~\um~--~160~\um.  {\color{black}While the shorter wavelengths observed by JWST (0.6~--~27.9~\um) do not correlate as well with luminosity, we demonstrate that using a single photometric measurement in two different JWST filters simultaneously can result in luminosity estimates that are less uncertain than even the best estimates obtained using a single JWST filter.  Using a single photometric measurement in three different JWST filters simultaneously can result in luminosity estimates that are comparable in accuracy to those obtained using single far-infrared photometric flux measurements.}
\end{abstract}




\color{black}
\section{Introduction}

Studies of the earliest stages of star formation, when a protostar is still actively accreting mass from its surroundings, are obscured by the surrounding, collapsing dense molecular cloud core of gas and dust \citep[e.g.,][]{difrancesco2007:ppv,wardthompson2007:ppv,enoch2009:protostars,evans2009:c2d,dunham2014:ppvi,fischer2023:ppvii}.  As a result, many of the properties of protostars that must be determined to fully understand the physical processes responsible for setting the final masses of stars are not directly observable, and must instead be inferred through indirect means.  To give one specific example, while the distribution of protostellar masses may be one of the best ways to distinguish between competing theoretical scenarios for the temporal evolution of mass accretion onto protostars \citep[e.g.,][]{mckee2010:masses,offner2011:luminosities,palau2024:review,hartmann2025:lbol-mass}, the only direct method to measure the mass of a protostar is via the detection and characterization of a Keplerian rotation disc.  While this has now become possible over the past decade with the very high sensitivity and spatial resolution provided by the Atacama Large Millimeter/submillimeter Array (ALMA), the required observations are very expensive in terms of observing time and thus the number of protostars with directly measured masses remains very small \citep[see][and references therein]{tobin2024:araa,palau2024:review}.

An indirect method of measuring both the masses and instantaneous accretion rates of protostars is by measuring their total bolometric luminosities (\lbol), which includes contributions from photospheric luminosity (\lstar), accretion luminosity (\lacc), and external heating (\lext):

\begin{equation}
    \lbol = \lstar + \lacc + \lext = \lint + \lext \qquad ,
\end{equation}
where the sum of the stellar and accretion luminosity components is often referred to as the internal luminosity, \lint, of a protostar.  Since the accretion luminosity depends on both the instantaneous mass of a protostar and its instantaneous mass accretion rate, the internal (and thus also the bolometric) luminosity can be used as an indirect tracer for both quantities.  

Indeed, a large body of work analyzing the bolometric luminosities of protostars has been published over the last several decades, dating back to a series of studies by \citet{kenyon1990:luminosities,kenyon1994:luminosities,kenyon1995:luminosities} that found protostars in Taurus to be underluminous compared to theoretical predictions.  This so-called ``luminosity problem'' has been extensively studied in the intervening decades, especially with large-area mid-infrared and far-infrared surveys conducted with the {\it Spitzer Space Telescope} ({\it Spitzer}) and {\it Herschel Space Observatory} ({\it Herschel}) \citep[e.g.,][]{young05:evolmodels,dunham2008:lowlum,enoch2009:protostars,evans2009:c2d,kryukova2012:luminosities,dunham2013:luminosities,fischer2017:luminosities}.  Based on these studies, our current understanding is that temporal accretion variability during the protostellar stage is required to match the observed luminosities of protostars, but there remains a degeneracy between viable theoretical accretion scenarios dominated by slow, secular accretion rate changes and those dominated by rapid, stochastic, large-amplitude accretion rate changes (the so-called ``episodic accretion'' scenario) \citep[e.g.,][]{dunham2014:ppvi,fischer2023:ppvii}.

One potential avenue to decipher the constraints that the observed luminosities of protostars place on the underlying accretion process is to vastly improve the statistical constraints provided by protostellar luminosities.  This would require moving beyond the relatively complete sample of $\sim$500 protostellar luminosities in nearby star-forming regions \citep[those within 500~pc of the Sun;][]{dunham2014:outflows} to the many thousands of protostars contained within larger and more distant star-forming regions.  However, accurately measuring the luminosity of a protostar requires sampling its spectral energy distribution (SED) over approximately three orders of magnitude in wavelength (between $\sim$1--1000~\um), with sampling at far-infrared and submillimeter wavelengths ($\sim$70--350~\um) especially important given that this is where protostellar SEDs typically peak \citep[see][for more details on the challenges in accurately measuring protostellar luminosities]{dunham2013:luminosities}.  Such work starts to become infeasible for star-forming regions more distant than 500~pc, as neither {\it Spitzer}, {\it Herschel}, nor the {\it Stratospheric Observatory for Infrared Astronomy (SOFIA)} achieve the sensitivity or angular resolution necessary to detect and resolve individual protostars at such distances over the full range of protostellar masses and luminosities.

As a work-around to the large, multi-wavelength datasets needed to accurately measure the luminosities of protostars, \citet{dunham2008:lowlum} (hereafter D08) examined correlations between \lint\ of a protostar and its flux $F_{\nu}$\footnote{The flux, $F_{\nu}$, is related to the flux density, $S_{\nu}$, as $F_{\nu} = \nu S_{\nu}$, where $F_{\nu}$ is measured in cgs units of erg~s$^{-1}$~cm$^{-2}$ and $S_{\nu}$ is measured in cgs units of erg~s$^{-1}$~cm$^{-2}$~Hz$^{-1}$.  While we follow standard notation and use $F_{70}$ (or $S_{70}$) to refer to the flux (or flux density) at the frequency corresponding to a wavelength of 70~\um, we emphasize that this flux (or flux density) does refer to $F_{\nu}$ (or $S_{\nu}$) and not $F_{\lambda}$ (or $S_{\lambda}$).} at the six wavelengths surveyed by {\it Spitzer} (3.6, 4.5, 5.8, 8.0, 24, and 70~\um).  They found that the observed fluxes at $3.6-24$~\um\ correlated poorly with \lint, but \fseventy, the observed flux at 70~\um, showed a tight linear correlation with \lint\ in log-log space, implying that a single photometric measurement of a protostar at 70~\um\ could be used to accurately estimate its luminosity.  These results were later confirmed by \citet{huard2017:luminosities} (hereafter HT17), who also examined the mid-infrared wavelength range of 19.7--37.1~\um\ observed by {\it SOFIA} \citep{adams2010:forcast,herter2012:forcast}.  HT17 found that fluxes at single {\it SOFIA} wavelengths correlate poorly with luminosity, consistent with previous results for similar {\it Spitzer} wavelengths, whereas combinations of fluxes measured at two SOFIA wavelengths can be used to estimate \lint\ with as much accuracy as the flux at 70~\um\ alone.  These results are broadly consistent with those of \citet{kryukova2012:luminosities}, who noted that a correlation between mid-infrared fluxes and protostellar luminosity is possible when the slope of the mid-infrared SED is also taken into account.  While the sensitivity and angular resolution of archival {\it Spitzer}, {\it SOFIA}, and {\it Herschel} data are insufficient to apply the results of D08 and HT17 to more distant star-forming regions, the {\it James Webb Space Telescope} (JWST) is currently offering a revolutionary increase in sensitivity and angular resolution at near and mid-infrared wavelengths, and is capable of both detecting and resolving individual protostars in star-forming regions well beyond those located within 500~pc of the Sun.

In this paper we revisit the extent to which the luminosities of protostars can be accurately estimated from a limited number of infrared measurements, using the evolutionary radiative transfer models presented by \citet{dunham2012:evolmodels}.  Our goals are three-fold: (1) confirm (and revise if necessary) the 
correlation between \fseventy\ and \lint\ found by both D08 and HT17, using radiative transfer models that span a larger range of parameters; (2) examine correlations between \fnu\ and \lint\ at \underline{\it all} wavelengths to better identify the best wavelength(s) to use; and (3) examine the extent to which JWST observations alone can be used to estimate the luminosities of protostars.  The organization of this paper is as follows: in Section \ref{sec_models} we describe the models used for this work.  In Section \ref{sec_results} we present our main analysis and results on estimating protostellar luminosities with infrared photometric measurements.  We begin this section with a short discussion of how we weight each model in our analysis (Section \ref{sec_results_weights}), followed by an analysis of the correlation between \fseventy\ and \lint\ in Section \ref{sec_results_f70}, an examination of the correlation between \fnu\ and \lint\ at all wavelengths in \S \ref{sec_results_allwv}, and then concluding with an investigation in Section \ref{sec_results_jwst} into the extent to which JWST observations can be used to estimate the luminosities of protostars.  We then summarize our findings in Section \ref{sec_summary}.

\section{Description of the Models}\label{sec_models}

To examine trends between the internal luminosities of protostars and their fluxes at various wavelengths, we use the evolutionary radiative transfer models presented by \citet{dunham2012:evolmodels}.  These models couple hydrodynamical simulations of dense cores collapsing to form protostars with two-dimensional, axisymmetric radiative transfer calculations to generate model SEDs at various inclination angles and timesteps throughout the collapse of each core.  While full details of the hydrodynamical simulations and radiative transfer calculations are presented by \citet{dunham2012:evolmodels}, we summarize the key details below.  We also emphasize here that we are using these specific models because they provide radiative transfer models over a wide range of core masses, protostellar masses, evolutionary stages, and inclination angles.

\subsection{Hydrodynamical Simulations}\label{sec_models_hydro}

\begin{table*}
\begin{center}
\caption{Initial Conditions For Each Simulation}
\label{tab_simulations}
\begin{tabular}{ccccccc}
\hline \hline
           & $\Omega_0$    & $r_0$ & $\Sigma_0$   & Core Outer Radius & Initial Core Mass &             \\
Simulation & (rad s$^{-1}$ & (au)  & (g cm$^{-2}$ & (pc)              & (\msun)           & $\beta$     \\
\hline
1           & 1.6\ee{-13}   & 685   & 0.18         & 0.02              & 0.305             & 8.75\ee{-3} \\
2           & 1.95\ee{-13}  & 685   & 0.18         & 0.02              & 0.305             & 1.26\ee{-2} \\
3           & 3.0\ee{-14}   & 3770  & 0.033        & 0.11              & 1.684             & 8.8\ee{-3}  \\
4$^{\rm a}$ & 2.5\ee{-14}   & ...   & 0.026        & 0.04              & 0.612             & 8.75\ee{-3} \\
5$^{\rm a}$ & 0.9\ee{-14}   & ...   & 0.0093       & 0.11              & 1.686             & 8.8\ee{-3}  \\
6$^{\rm a}$ & 1.2\ee{-14}   & ...   & 0.013        & 0.08              & 1.22              & 8.8\ee{-3}  \\
7           & 2.0\ee{-13}   & 445   & 0.28         & 0.013             & 0.194             & 5.6\ee{-3}  \\
8           & 2.3\ee{-14}   & 3770  & 0.033        & 0.11              & 1.689             & 5.6\ee{-3}  \\
9           & 2.6\ee{-13}   & 514   & 0.24         & 0.015             & 0.229             & 1.26\ee{-2} \\
10          & 2.9\ee{-14}   & 3085  & 0.04         & 0.09              & 1.378             & 5.6\ee{-3}  \\
11          & 3.2\ee{-14}   & 4115  & 0.03         & 0.12              & 1.84              & 1.26\ee{-2} \\
12          & 3.7\ee{-14}   & 2400  & 0.05         & 0.07              & 1.0726            & 5.6\ee{-3}  \\
13          & 1.6\ee{-13}   & 685   & 0.18         & 0.02              & 0.305             & 8.75\ee{-3} \\
14          & 3.9\ee{-13}   & 342   & 0.36         & 0.01              & 0.1515            & 1.26\ee{-2} \\
15          & 3.9\ee{-14}   & 3430  & 0.036        & 0.1               & 1.5312            & 1.26\ee{-2} \\
16          & 3.25\ee{-13}  & 274   & 0.45         & 0.008             & 0.1174            & 5.6\ee{-3}  \\
17          & 4.7\ee{-14}   & 1885  & 0.066        & 0.055             & 0.8434            & 5.6\ee{-3}  \\
18          & 4.8\ee{-14}   & 2745  & 0.045        & 0.08              & 1.2422            & 1.26\ee{-2} \\
19          & 5.56\ee{-13}  & 240   & 0.52         & 0.007             & 0.1052            & 1.26\ee{-2} \\
20          & 1.1\ee{-13}   & 1200  & 0.1          & 0.035             & 0.5349            & 1.26\ee{-2} \\
21          & 6.0\ee{-14}   & 1370  & 0.09         & 0.04              & 0.6084            & 5.6\ee{-3}  \\
22          & 2.8\ee{-14}   & 2230  & 0.056        & 0.065             & 0.999             & 2.75\ee{-3} \\
23          & 2.0\ee{-14}   & 2915  & 0.043        & 0.085             & 1.304             & 2.75\ee{-3} \\
\hline 
\end{tabular}
\end{center}
$^{\rm a}$Simulation featuring constant surface density and angular velocity profiles.\\
\end{table*}

The hydrodynamical simulations, first described in \citet{vorobyov2005:bursts, vorobyov2006:bursts, vorobyov2010:bursts}, calculate the collapse of gravitationally bound dense cores.  The simulations begin in the prestellar phase and utilize a 6~au sink cell into which material can enter but not exit.  This sink cell acts as a gravitating point-mass star, and is introduced when the density in the sink exceeds a threshold value of 10$^{11}$~cm$^{-3}$. Subsequently, a circumstellar disk forms around the sink when the spinning-up inner layers of the core hit a centrifugal barrier near the sink.  The disks in these simulations include the effects of stellar and background irradiation, viscous and shock heating, self-gravity within the disc, disc radiative cooling, and mass-loss in protostellar jets (the latter introduced phenomenologically by assuming that 10\% of the mass that passes through the sink is carried away by the jets rather than landing on the protostar).  These are two-dimensional simulations that operate in the thin-disc limit so that non-axisymmetric structure can be modeled while maintaining sufficient computational simplicity to follow the full growth of the protostar from its formation through to its final mass.  Although not relevant for this specific study, we note here that these specific simulations feature significant axial asymmetries and predict a stochastic accretion process with short-duration, large-amplitude bursts \citep{vorobyov2005:bursts,vorobyov2006:bursts,vorobyov2009:bursts,vorobyov2010:bursts,vorobyov2015:bursts}.

A total of 23 simulation runs spanning a large range of initial conditions were considered by \citet{dunham2012:evolmodels}.  All simulations start with a constant initial core gas temperature of 10~K.  Most simulations start with gas surface density $\Sigma$ and angular velocity $\Omega$ profiles that scale as $\Sigma \propto r^{-1}$ and $\Omega \propto r^{-1}$.  The specific profiles, which are typical for magnetically supercritical cores formed by slow gravitational contraction \citep{Basu1997}, are given by:

\begin{equation}
    \Sigma(r) = \frac{r_0 \Sigma_0}{\sqrt{r^2 + r_0^2}} \qquad ,
\end{equation}

\begin{equation}
    \Omega(r) = 2 \Omega_0 \left( \frac{r_0}{r} \right)^2 \left[\sqrt{1 + \left( \frac{r}{r_0}^2 \right)} -1   \right] \qquad ,
\end{equation}

\noindent
where $r_0$ is the radius of the central region with near-constant surface density, $\Omega_0$ is the central angular velocity, and $\Sigma_0$ is the central surface density.  Each simulation is truncated so that the ratio of the core radius to the radius of the central flat region ($r_{\rm core}/r_0$) is constant and equal to 6.0.  Finally, as an alternative, three simulations adopt radially constant surface density and angular velocity profiles, where $\Sigma(r)~=~\Sigma_0$ and $\Omega(r)~=~\Omega_0$ for all radii.  Table \ref{tab_simulations} lists $\Omega_0$, $r_0$, and $\Sigma_0$ for all 23 simulation runs, and indicates the three simulations with radially constant profiles.  Also listed in Table \ref{tab_simulations} are the core outer radius, initial core mass, and initial ratio of rotational to gravitational energy ($\beta$).  Additional details on these parameters, including a discussion of their values in comparison to the observed properties of dense cores, can be found in \citet{dunham2012:evolmodels}, as well as \citet{vorobyov2005:bursts, vorobyov2006:bursts, vorobyov2010:bursts}.  


\subsection{Evolutionary Radiative Transfer Models}\label{sec_models_radtrans}

\citet{dunham2012:evolmodels} coupled the hydrodynamical simulations described above with radiative transfer calculations at various timesteps throughout each simulation run, constructing what they called ``evolutionary radiative transfer models'' to predict observable quantities (more specifically, the spectral energy distribution at each timestep).  While an ideal coupling would take the detailed disc and core density structures directly from the simulations and input them into the radiative transfer calculations, the two-dimensional simulations were conducted in the thin-disc approximation, with only the cylindrical (midplane) radius and azimuthal angle specified.  On the other hand, the radiative transfer package available at the time was {\sc RADMC}, a two-dimensional, axisymmetric, Monte Carlo dust radiative transfer package \citep{dullemond2000:radmc,dullemond2004:radmc}, with only the spherical radius and polar angle specified.  Thus the simulations did not specify the full vertical structure of the collapsing protostellar systems, and the radiative transfer calculations did not incorporate the axial asymmetries featured in the simulations.  To address this limitation, \citet{dunham2012:evolmodels} used analytic profiles for the core and disc density structures, scaled to match the physical parameters from the simulations.  More specifically, the simulations provide the initial core mass (\mcore) and outer radius\footnote{Following the convention adopted by \citet{dunham2012:evolmodels}, a lowercase $r$ is used for radii pertaining to the core and uppercase $R$ is used for radii pertaining to the protostar and disc.} (\rcore), along with the time evolution of the core mass, disc mass (\mdisc), protostellar mass (\mstar), disc outer radius (\rdisc), accretion rate onto the protostar (\mdotstar), and accretion rate onto the disc (\mdotdisc).  While we use the first-person ``we'' to describe the setup of these evolutionary radiative transfer models in the rest of this subsection, we emphasize here that we are simply summarizing the work presented by \citet{dunham2012:evolmodels}.

Since the simulations begin in the prestellar phase and extend well beyond the end of the embedded protostellar phase, we select only the subset of timesteps in each simulation that begin with the formation of the first hydrostatic core \citep[FHSC;][]{larson1969:fhsc} and end when the instantaneous core mass declines to 10\% of the initial core mass \citep[which is the moment we define as the end of the embedded stage; see][for details]{vorobyov2009:bursts,vorobyov2010:bursts}.  We also resampled the simulation timesteps, which number in the thousands to tens of thousands per simulation due to the very high time resolution of the simulations, to a much coarser grid of timesteps due to technical limitations in the number of timesteps feasible for the radiative transfer calculations.  This resampling started with a regular grid of up to $\sim$500 timesteps but was then modified to ensure high-amplitude accretion bursts were preserved \citep[see Appendix~A of][for details]{dunham2012:evolmodels}.  Of relevance for this work is the fact that the resampled grid of timesteps does not feature a constant interval between timesteps, as described in more detail in \S~\ref{sec_results_f70} below.

At each resampled timestep of each model, the disc density profile is assumed to have a power-law radial dependence and Gaussian vertical dependence, given as:

\begin{equation}
    \rho_{\rm disc}(s,z) = \rho_0 \left( \frac{s}{s_0}\right)^{-\alpha} {\rm exp} \left[ -\frac{1}{2} \left( \frac{z}{H_s} \right)^2\right] \qquad ,
\end{equation}

\noindent where $z$ is the vertical coordinate (distance above the midplane; $z=r\cos{\theta}$, with $r$ and $\theta$ the radial and zenith angle spherical coordinates), $s$ is the midplane radius ($s = \sqrt{r^2 - z^2}$), $H_s$ is the disc scale height \citep[described in more detail in ][]{dunham2012:evolmodels}, and $\rho_0$ is the density in the midplane at the reference coordinate $s_0$.  This density profile is truncated at the outer radius \rdisc\ specified by the simulations, and at the inner radius ($R_{\rm disc}^{\rm in}$) where the dust temperature reaches the dust sublimation temperature.  This inner radius is calculated, assuming spherical dust grains, as

\begin{equation}
    R_{\rm disc}^{\rm in} = \sqrt{\frac{L_*}{16 \pi \sigma T_{\rm dust}^4}} \qquad ,
\end{equation}

\noindent where $L_*$ is the protostellar luminosity (calculated as described below) and $T_{\rm dust}$ is the assumed dust sublimation temperature of 1500~K \citep[e.g., ][]{cieza2005:tdust}.  After truncation at the inner and outer radii, the parameter $\rho_0$ is set so that \mdisc\ matches that from the simulations.

The core density profile at each resampled timestep is given by the \citet[][hereafter TSC84]{terebey1984:model} solution for the collapse of a slowly rotating core, which results in a density profile that is initially a spherically symmetric, singular isothermal sphere identical to the non-rotating case considered by \citet{shu1977:sis}.  As collapse proceeds, this density profile evolves to take on two forms: an outer solution identical to the non-rotating case and an inner solution that exhibits rotational flattening, with the transition radius between the two moving outward with time.  We truncate the core density profile at each resampled timestep at the given outer radius (\rcore) given by the simulations, and at an inner radius equal to the disc outer radius (\rdisc)\footnote{At early times when the disc is very small this results in unreasonably large optical depths through the core, thus we adopt a minimum core inner radius such that the initial optical depth at 100 \um\ does not exceed a value of 10; see \citet{dunham2010:evolmodels,dunham2012:evolmodels} for details.}.  At late times we allow \rcore\ to decrease following the collapse velocities given by the TSC84 solution \citep[see][for details]{dunham2010:evolmodels,dunham2012:evolmodels}.  We normalize the density profile at each resampled timestep so that \mcore\ matches that from the simulations.  While this renormalization means that the TSC84 solution is no longer self-consistent from one timestep to the next, \citet{dunham2010:evolmodels} and \citet{dunham2012:evolmodels} made this choice to preserve both the exact evolution of core mass predicted by the simulations and the qualitative feature of the TSC84 solution that the effects of rotational flattening grow more significant with time, and are overall more significant in cores with larger initial angular velocities.

The radiative transfer calculations include both internal and external luminosity sources.  The external luminosity, which is treated as an isotropic radiation field incident on the outer radius of the core, arises from the interstellar radiation field (ISRF).  Following the approach originally developed by \citet{evans2001:starless} \citep[see also][]{shirley2002:scuba,young05:evolmodels}, we adopt the \citet{black1994:isrf} ISRF, modified in the ultraviolet to reproduce the \citet{draine1978:isrf} ISRF, and then extincted by $A_V = 0.5$ of dust with properties given by \citet{draine1984:dust} to simulate extinction by the surrounding parent molecular cloud.

The internal luminosity is treated as a point source of radiation at the center of the model.  This luminosity, consisting of the total luminosity of the protostar and disc, consists of six separate components: (1) accretion luminosity from mass accretion directly from the core onto the protostar, (2) accretion luminosity from mass accretion from the core onto the disc, (3) accretion luminosity from mass accretion from the disc onto the protostar, (4) "mixing luminosity" from the mixing of mass newly accreted onto the disc with existing disc material, (5) luminosity due to the release of energy stored in differential rotation of the protostar, and (6) protostellar photosphere luminosity due to gravitational contraction and deuterium burning.  Specific details on how these components are calculated using the simulation outputs are described in \citet{dunham2012:evolmodels} \citep[see also ][]{adams1986:protostars,young05:evolmodels,dunham2010:evolmodels}.

For each resampled timestep of each simulation, we use {\sc RADMC} to calculate the two-dimensional dust temperature profile of the core.  
The dust opacities are those given by \citet{ossenkopf1994:oh5} appropriate for thin ice mantles after $10^5$~yr of coagulation at a gas density of $10^6$~cm$^{-3}$ (OH5 dust), which previous studies have shown are appropriate for cold, dense cores \citep[e.g.,][]{evans2001:starless,shirley2005:l1498}, {\color{black}modified to include isotropic scattering as described in \citet{young05:evolmodels} and \citet{dunham2010:evolmodels} (see \S \ref{sec_limitations_scattering} below for further discussion of scattering). } 
{\sc RADMC} is then used to calculate SEDs at nine different inclinations (5\degree\ to 85\degree, in steps of 10\degree, where $i=0$\degree\ corresponds to a face-on system and $i=90\degree$ corresponds to an edge-on system).

\begin{figure*}
    \resizebox{15cm}{!}{\includegraphics{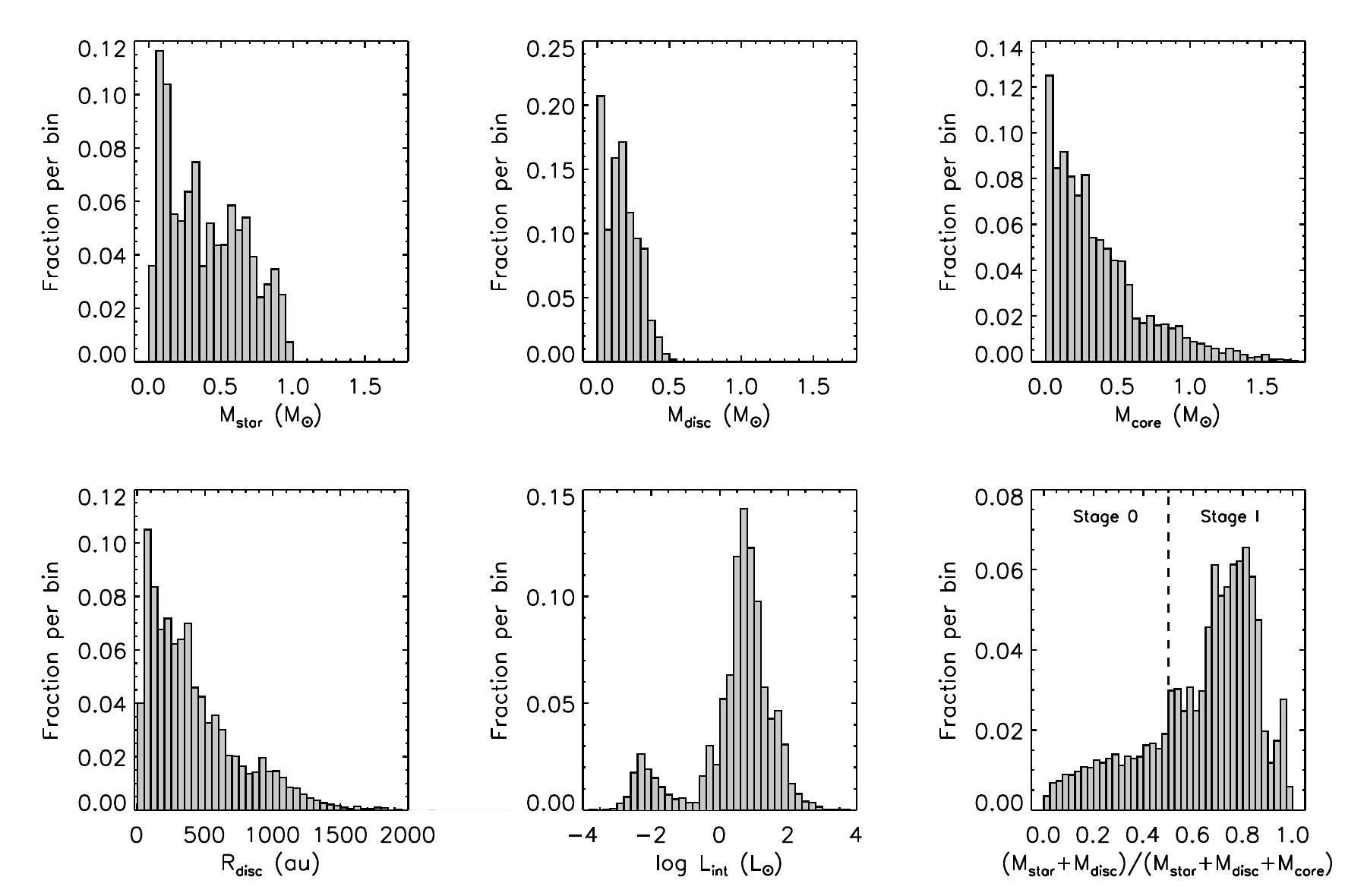}}
    \vspace{0.0in}
    \caption{Histograms showing the distribution of instantaneous (current) physical parameters in the 9,990 radiative transfer models used in this study, including (left-to-right, top-to-bottom) instantaneous protostellar mass (\mstar), disc mass (\mdisc), core mass (\mcore), disc radius (\rdisc), internal luminosity (\lint), and ratio of internal $(\mstar + \mdisc)$ to total $(\mstar + \mdisc + \mcore)$ mass.  The dip in the bottom center panel near $\log(\lint) \approx 0$ is due to the rapid transition between first and second hydrostatic cores in the simulations.  The dashed vertical line in the last panel shows the boundary between Stage 0 (younger) and Stage I (older) protostars, defined as the point at which half of the total model mass has accreted onto the star+disc system.}
\label{fig_model_parameters}
\end{figure*}

With 23 simulation runs, each sampled at several hundred timesteps (generally between approximately $100-600$ timesteps per simulation run) and each timestep analyzed at nine different inclinations, we have a total of 9,990 physical radiative transfer models generating a total of 89,910 SEDs.  In the remainder of this paper, we refer to these 89,910 model SEDs as the "models."  Figure \ref{fig_model_parameters}, which plots histograms showing the range and distribution of protostellar masses (\mstar), disc masses (\mdisc), core masses (\mcore), disc radii (\rdisc), internal luminosities (\lint), and ratio of internal $(\mstar + \mdisc)$ to total $(\mstar + \mdisc + \mcore)$ mass (a measure of relative evolutionary status), demonstrates that these models span a very large range of basic physical parameters.  For the remainder of this work we use these evolutionary radiative transfer models as a large database of model SEDs; the underlying physical nature of the simulations themselves are not relevant to this current study.

\begin{figure}
    \resizebox{\hsize}{!}{\includegraphics{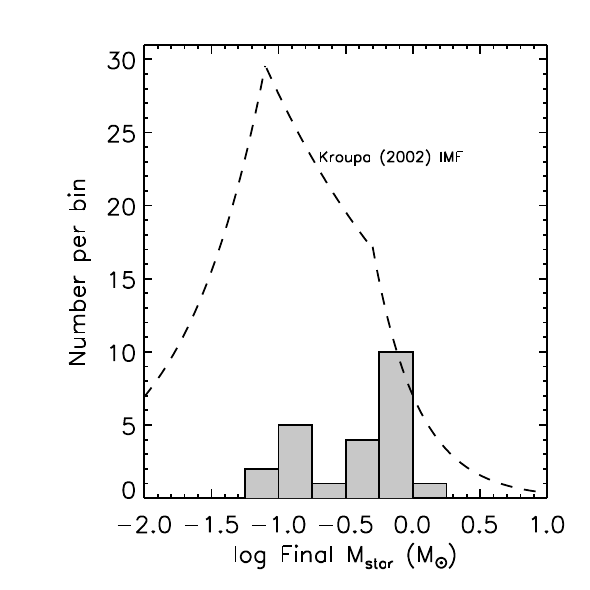}}
    \vspace{0.0in}
    \caption{Histogram showing the distribution of the final masses of the stars formed in the 23 simulation runs used in this study.  The dashed line shows the \citet{kroupa2002:imf} three-component power-law IMF, normalized to match the number of simulations forming stars with masses in the bin centered at $\log(\rm final\, \mstar/\msun)~=~-0.125 $.  Note the linear rather than logarithmic scaling of the vertical axis.}
\label{fig_mstar_final}
\end{figure}

\section{Estimating the Luminosities of Protostars}\label{sec_results}

\subsection{Assigning a weight to each model}\label{sec_results_weights}

Before analyzing how well the fluxes at various wavelengths (or combinations of wavelengths) correlate with the internal luminosity of the protostar, we must first consider three points about the models:

\begin{enumerate}
    \item The 23 simulation runs form stars with final masses ranging from 0.077~\msun\ to 1.0~\msun\ ($81-1048$~M$_{\rm Jupiter}$), with a distribution that is not representative of the stellar initial mass function (IMF).  Indeed, as shown in Figure \ref{fig_mstar_final}, simulations forming stars with $M~\sim$~1~\msun\ are significantly over-represented relative to those forming stars with lower masses when compared to the \citet{kroupa2002:imf} three-component power-law IMF.  The simulations also only sample the very top end of the substellar mass regime, as they were designed specifically to simulate the formation of low-mass stars.
    
    \item With SEDs generated at inclinations ranging from 5\degree\ to 85\degree, in steps of 10\degree, we have just as many low inclination (nearly face-on) models as high inclination (nearly edge-on) models.  In reality, high inclination models are much more likely to be observed due to the increased solid angle at high inclinations compared to low inclinations.
    
    \item The resampled timesteps use a much smaller time interval during bursts of high accretion (and thus high luminosity) compared to all other times, since temporally resolving the accretion bursts was an important component of the work presented by \citet{dunham2012:evolmodels}.  As a consequence, we have a much larger fraction of model SEDs undergoing accretion bursts (and thus at very high luminosities) than the actual fractions of such protostars.
\end{enumerate}

The result of the above points is that models forming higher mass stars, models viewed at lower inclinations, and models with very high luminosities are all overrepresented compared to the actual distribution of observed protostars in any large, statistically robust sample.  To ensure these rare models do not bias the various fits we will perform, we first assign a weight to each model, $w_{\rm total}$, defined as the product of three normalized weights:

\begin{equation}\label{eq_weights}
    w_{\rm total} = w_{\rm IMF} w_{\rm inc} w_{\rm time} \qquad .
\end{equation}

In Equation \ref{eq_weights}, we follow \citet{dunham2012:evolmodels} and set $w_{\rm IMF}$ for each model based on the final stellar mass produced by the simulation run from which the model is taken.  We adopt the \citet{kroupa2002:imf} three-component power-law IMF, which gives $dN/dM~\propto~M^{-\alpha}$, with $\alpha~=~0.3$ for $0.01~\leq~M/\msun~<0.08$, $\alpha~=~1.3$ for $0.08~\leq~M/\msun~<0.5$, and $\alpha~=~2.3$ for $M/\msun~\geq~0.5$.  We first assign $w_{\rm IMF}$ to each model using an arbitrary normalization for the \citet{kroupa2002:imf} IMF, and then normalize these weights so that the maximum value of $w_{\rm IMF}$ is equal to one.

Following \citet{dunham2010:evolmodels} and \citet{dunham2012:evolmodels}, we first set $w_{\rm inc}$ for each model equal to the fraction of the total solid angle subtended by the inclination of that model, calculated in practice by assuming model SEDs at each of the nine inclinations considered ($i~=$ 5\degree, 15\degree, 25\degree, ..., 85\degree) are valid for inclinations spanning from $(i~-~5\degree)$ to $(i~+~5\degree)$.  We then normalize these weights so that the maximum value of $w_{\rm inc}$ is equal to one.  Finally, we initially set $w_{\rm time}$ for each model equal to the total time (in kyr) encompassed by that model's resampled timestep, and then normalize these weights to that the maximum value of $w_{\rm time}$ is equal to one.

\subsection{Correlation between luminosity and \fseventy}\label{sec_results_f70}

\begin{figure}
    \resizebox{\hsize}{!}{\includegraphics{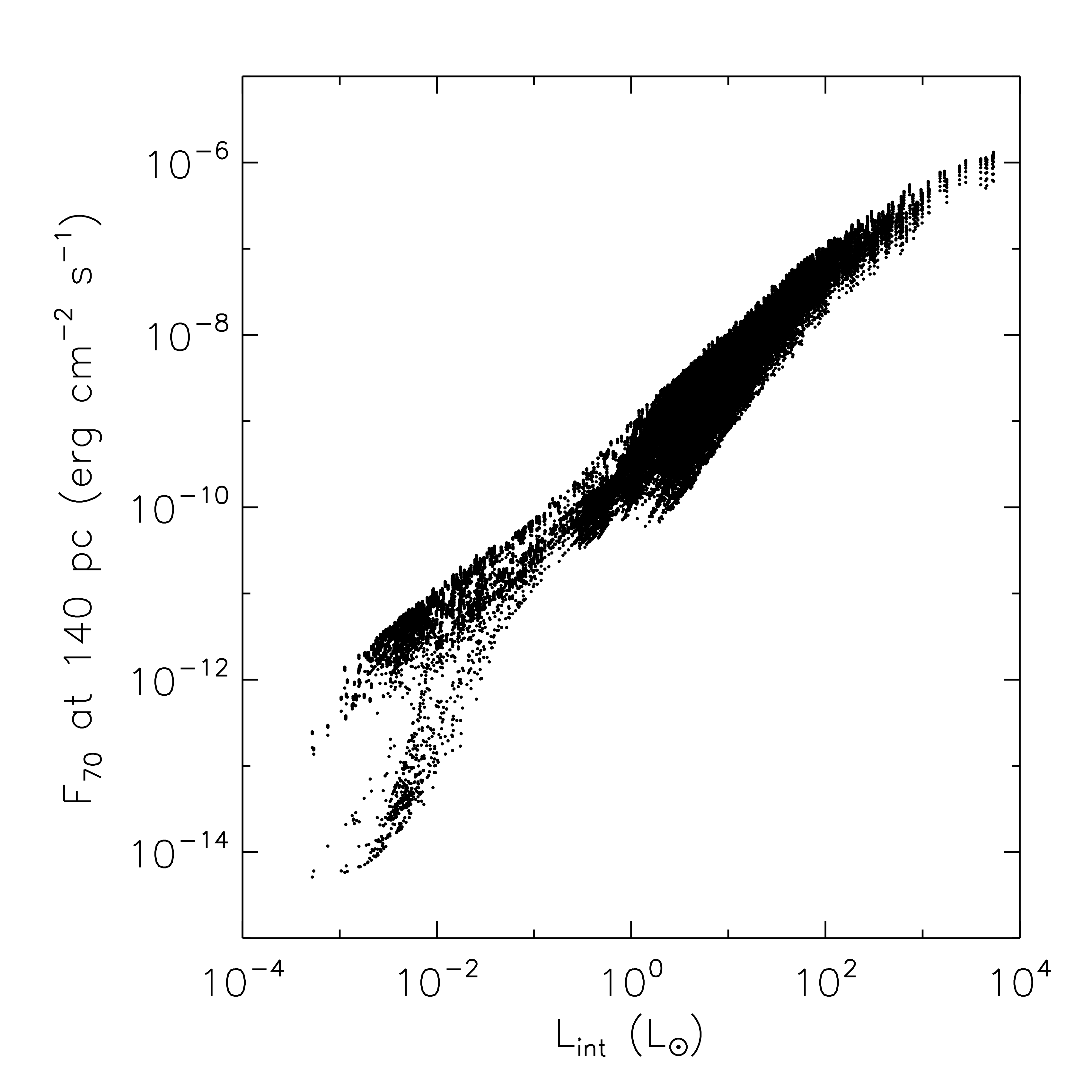}}
    \vspace{0.0in}
    \caption{\fseventy\ vs.~\lint\ from the 89,910 SEDs generated for our radiative transfer models.  The lower branch of points at low luminosities are primarily high inclination ($i=85\degree$) models with very low fluxes as a result of the very high inclinations.}
\label{fig_f70all}
\end{figure}

Figure \ref{fig_f70all} plots \fseventy\ vs.~\lint\ for the full set of 89,910 models, where the same symbol size is used for each model regardless of that model's weight.  \fseventy\ is determined from the model SEDs and \lint\ is the combined luminosity of the star and disc (the ``internal'' luminosity sources), and is calculated as the sum of the six internal luminosity components described above in \S \ref{sec_models_radtrans}.  We re-scale the SEDs generated by {\sc RADMC} from their default distance of 1 pc to $d~=~140$~pc, and we use linear interpolation in log-log space to calculate the flux at \fseventy\ (since the closest wavelengths in the wavelength grid used by {\sc RADMC} are 63.1~\um\ and 73.6~\um).  

Following D08 and HT17, we perform a linear least-squares fit to $\log (\fseventy)$ vs.~$\log (\lint)$.  Specifically, we fit the following function:

\begin{equation}\label{eq_fit_f70}
    \log (\fseventy) = c_0 + c_1 \log (\lint) \qquad ,
\end{equation}

\noindent with \lint\ in units of \lsun\ and \fseventy\ in units of erg~s$^{-1}$~cm$^{-2}$, normalized to its value at an assumed distance of 140~pc. Although we ultimately desire to use this fit to estimate \lint\ from observed values of \fseventy, D08 and HT17 fit $\log (\fseventy)$ vs.~$\log (\lint)$ (rather than the other way around) because it is the internal luminosity of the protostar that determines its flux at 70~\um, and thus it is indeed most appropriate to treat \lint\ as the independent variable.

\begin{figure*}
    \resizebox{7.0in}{!}{\includegraphics{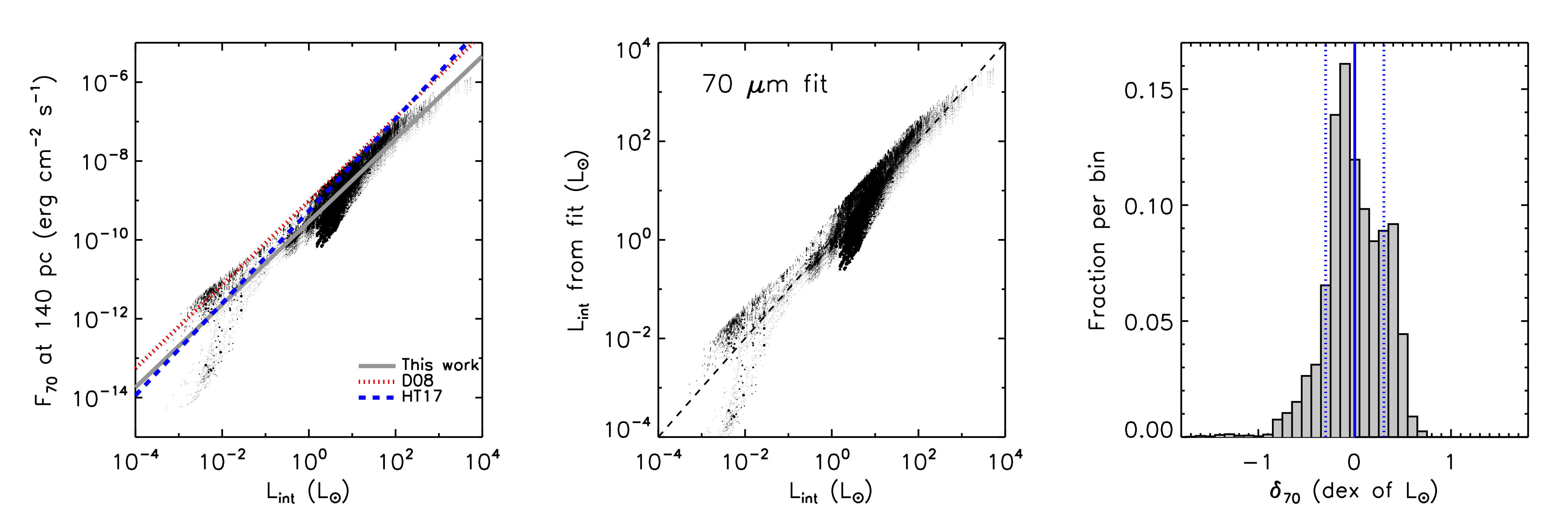}}
    \vspace{0.0in}
    \caption{\underline{Left:} \fseventy\ vs.~\lint\ for the 89,910 SEDs generated from our radiative transfer models.  The symbol size for each model is proportional to the value of $w_{\rm total}$ for that model, where $w_{\rm total}$ is calculated as described in \S \ref{sec_results_weights}.  The thick gray line shows the linear least-squares fit in log-log space from this current study, while the dotted red line and dashed blue line show the same from D08 and HT17, respectively.  \underline{Center:}  \lint\ estimated from the best-fit to Equation \ref{eq_fit_f70} using the model value of \fseventy\ plotted vs.~the intrinsic \lint\ in the model. The symbol size for each point is proportional to the value of $w_{\rm total}$ for that model, where $w_{\rm total}$ is calculated as described in \S \ref{sec_results_weights}.  The dashed line shows the one-to-one line.  \underline{Right:} Weighted histogram of $\delta$ values for the $\log(\fseventy)$~vs.~$\log(\lint)$ fit in this work, where $\delta$ is calculated as defined in Equation \ref{eq_delta} and each individual value of $\delta$ is weighted in the histogram by the $w_{\rm total}$ associated with the model that produced that $\delta$.  The solid blue vertical line indicates the mean value, and the dotted blue vertical lines indicate $\pm~1\sigma_{\rm \delta}$ from the mean.}
\label{fig_f70all_fit}
\end{figure*}

Unlike D08 and HT17, who both perform unweighted fits, in this work we weight each point in the fit by its value of $w_{\rm total}$ to account for the different likelihoods of each underlying model being observed.  The best-fit values of $c_0$ and $c_1$ are reported in the first row of Table \ref{tab_f70}, along with those from D08 and HT17 (in the second and third rows) for ease of comparison.  The left panel of Figure \ref{fig_f70all_fit} again plots \fseventy\ vs.~\lint, similar to Figure \ref{fig_f70all}, except with symbol sizes proportional to $w_{\rm total}$ for each model, and with the best-fit line (in log-log space) overplotted.  The best-fit lines from D08 and HT17 are also overplotted for comparison.  If we define $L_{\rm int, fit}$ as the result from solving Equation \ref{eq_fit_f70} for \lint\ and inserting the observed value of \fseventy\ from the model SEDs produced by {\sc RADMC}, and $L_{\rm int, 0}$ as the intrinsic value of the internal luminosity (given as an input to the radiative transfer model), the center panel of Figure \ref{fig_f70all_fit} plots $L_{\rm int, 0}$ vs.~$L_{\rm int, fit}$.

\begin{table*}
\begin{center}
\caption{Results of linear least-squares fits to $\log{\fseventy}$ vs.~$\log{\lint}$}
\label{tab_f70}
\begin{tabular}{llccc}
\hline \hline
Study     & Stage                      & $c_0 \pm \Delta~c_0$ & $c_1 \pm \Delta~c_1$ & $\sigma_{\delta}$ \\
\hline
This work & All (Stage 0 + Stage I)    & $-9.541~\pm~0.001$   & $1.050~\pm~0.001$    & 0.302             \\
HT17      & Stage 0$^{\rm a}$          & $-9.270~\pm~0.002$   & $1.169~\pm~0.003$    & 0.12$^{\rm b}$    \\
D08       & Stage 0$^{\rm a}$          & $-9.02~\pm~0.01$     & $1.06~\pm~0.01$      & ...$^{\rm c}$     \\
This work & Stage 0                    & $-9.261~\pm~0.002$   & $1.085~\pm~0.002$    & 0.249             \\
This work & Stage I                    & $-9.616~\pm~0.001$   & $1.040~\pm~0.002$    & 0.267             \\
\hline 
\end{tabular}
\end{center}
$^{\rm a}$While HT17 and D08 did not explicitly restrict their fits to Stage~0 models, their models mostly or entirely encompassed Stage~0 protostars only.\\
$^{\rm b}$This quantity was referred to as $\sigma_L$ in HT17.\\
$^{\rm c}$D08 did not calculate this quantity.\\
\end{table*}

To assess the reliability of our fit to $\log(\fseventy)$~vs.~$\log(\lint)$, in particular the typical uncertainty introduced when using this fit to estimate the luminosity of a protostar from its observed 70~\um\ flux, we define the fit deviation $\delta_{70}$ for each model as:

\begin{equation}\label{eq_delta}
    \delta_{70} = \log (L_{\rm int, fit}) - \log (L_{\rm int, 0}) \qquad .
\end{equation}

\noindent We then construct a weighted histogram of $\delta_{70}$, weighting each value by the $w_{\rm total}$ of the corresponding model to ensure that less commonly occurring models have a corresponding smaller impact on the distribution of $\delta_{70}$ values.

The right panel of Figure \ref{fig_f70all_fit} plots the weighted histogram of $\delta_{70}$ (in units of dex of solar luminosity).  The mean value of this weighted histogram is 0.0004~$\pm$~0.0007 dex.  The standard deviation is $\sigma_{\delta_{70}}~=~0.302$ dex, as reported in Table \ref{tab_f70} (we note here that this is very similar to the value defined as $\sigma_L$ in HT17, with the primary difference being our use of a weighted distribution).  These mean and standard deviation values indicate that, while our fit gives an estimated \lint\ that is correct on average, the 1$\sigma$ scatter between the estimated and true luminosities is 0.302 dex in \lsun, or a factor of $10^{0.302}~=~2.0$.  In other words, protostellar internal luminosities estimated from this method are correct {\it on average}, but on a case-by-case basis they have a 1$\sigma$ uncertainty of a factor of two.

It is worth noting that our value of $\sigma_{\delta_{70}}~=~0.302$ for the $\log (\fseventy)$ vs.~$\log (\lint)$ linear least-squares fit is about a factor of 2.5 larger than the value of $\sigma_{L}~=~0.12$ reported by HT17.  Since our models span a much larger range of parameters than those presented by HT17, we recalculated $\sigma_{\delta_{70}}$ using only models with $0.03~\leq~\lint/\lsun~\leq~30$ and $1.0~\leq~\mcore/\msun~\leq~10.0$ {\color{black} to match the ranges of \lint\ and \mcore\ considered by HT17.  Restricting to only these models gives $\sigma_{\delta_{70}}~=~0.15$, consistent with HT17's value of $\sigma_{L}~=~0.12$.  Furthermore, imposing these restrictions gives a fit with $c_0=-9.228 \pm 0.006$ and $c_1=1.101 \pm 0.008$, both of which are closer to the HT17 values than our fit to all models (see Table \ref{tab_f70}).  While these findings suggest that our increased scatter and different fit values compared to HT17 can largely be explained by the larger range of parameters considered here, we caution that imposing these restrictions only leaves us with 2070 models (2.3\% of our total models).  The vast majority of our models feature $\mcore < 1.0\ \msun$, as seen in Figure \ref{fig_model_parameters}.  This lack of higher-mass models is briefly discussed in \S \ref{sec_limitations_mcore}.}  

\begin{figure}
    \resizebox{3.0in}{!}{\includegraphics{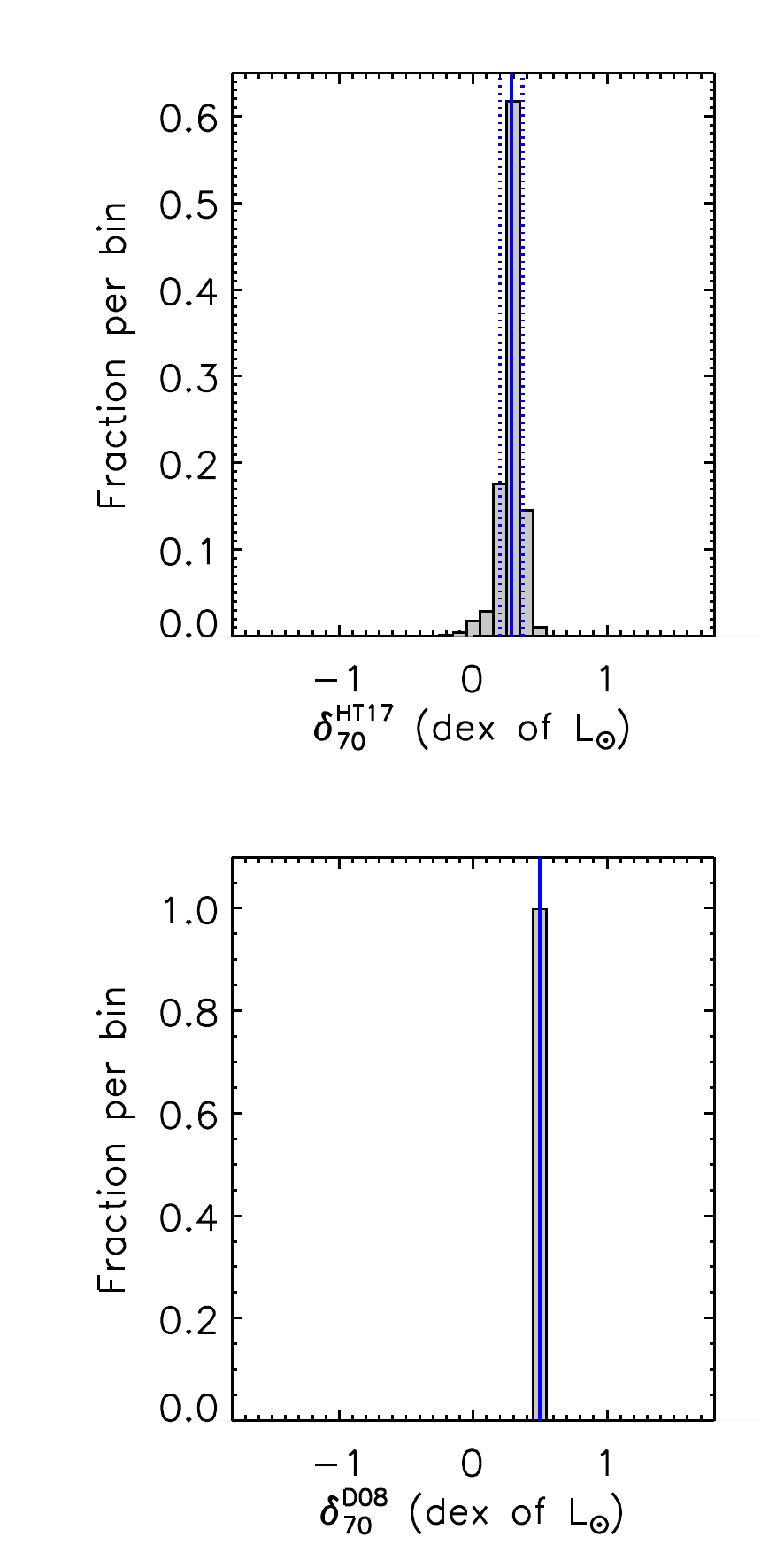}}
    \vspace{0.0in}
    \caption{Weighted histogram of $\delta_{70}^{\rm HT17}$ (top) and $\delta_{70}^{\rm D08}$ (bottom) values for the $\log(\fseventy)$~vs.~$\log(\lint)$ fit in this work, calculated as defined in Equations \ref{eq_delta_HT17} and \ref{eq_delta_D08}, with each individual value weighted by the $w_{\rm total}$ associated with the corresponding model.  The solid blue vertical line indicates the mean value, and the dotted blue vertical lines indicate $\pm1$ standard deviation from the mean.  The values of $\delta_{70}^{\rm D08}$ are tightly clustered around a single value because the D08 fit features a very similar slope but different intercept.}
\label{fig_f70all_deviatioins_others}
\end{figure}

To quantitatively compare our fit results to those of HT17 and D08, we calculate fit deviations between the estimated luminosity determined using our fit and that determined using either the HT17 or D08 fit as follows:

\begin{align}
    \label{eq_delta_HT17} \delta_{70}^{\rm HT17} &= \log (L_{\rm int, fit}) - \log (L_{\rm int, fit-HT17}) \\
    \label{eq_delta_D08} \delta_{70}^{\rm D08} &= \log (L_{\rm int, fit}) - \log (L_{\rm int, fit-D08}) \qquad ,
\end{align}

\noindent where $L_{\rm int, fit-HT17}$ is the estimated luminosity determined from \fseventy\ using the linear least-squares fit from HT17, and $L_{\rm int, fit-D08}$ is the estimated luminosity determined from \fseventy\ using the linear least-squares fit from D08.  The weighted distributions of $\delta_{70}^{\rm HT17}$ and $\delta_{70}^{\rm D08}$ are shown in Figure \ref{fig_f70all_deviatioins_others}, where the weighting is the same as described previously.  The mean and standard deviation of $\delta_{70}^{\rm HT17}$ are 0.286 and 0.085 dex of \lsun, respectively, indicating that using the HT17 fit will provide luminosity estimates that are, on average, a factor of $10^{0.286}~=~1.9$ too low.  The mean and standard deviation of $\delta_{70}^{\rm D08}$ are 0.500 and 0.001 dex of \lsun, respectively, indicating that using the D08 fit will provide luminosity estimates that are, on average, a factor of $10^{0.500}~=~3.2$ too low.

\subsubsection{Separating into evolutionary stages}\label{sec_results_f70_2stage}

The protostellar stage is typically divided into two observational classes, called Class~0 and Class~I.  Class~0 protostars were originally defined by \citet{andre1993:class0} to be the youngest protostars, specifically those that have accreted less than half of their final mass, whereas Class~I protostars are those that have accreted greater than half of their final mass.  Since the current mass of a protostar is difficult to measure (the only direct measurement method is based on the detection of Keplerian rotation in the surrounding disc, requiring (sub)millimeter wavelength molecular line observations with very high sensitivity and angular resolution), in practice protostellar classification is typically performed using the shape of the observed SED \citep[see, e.g., ][]{andre1993:class0,myers1993:tbol,evans2009:c2d,dunham2014:ppvi,frimann2016:simulations}.  Here we follow the convention introduced by \citet{robitaille2006:models} where `Class' is determined by observed characteristics (often the shape of the SED) and `Stage' is determined by physical characteristics, with Stage~0 protostars referring to those that have accreted less than half of their final mass, and Stage~I protostars referring to those have accreted half or more of their final mass.  It is assumed that there is a one-to-one correspondence between Class and Stage, although in practice that is not always possible to confirm.

Given that the models used by HT17 adopted a fix protostellar mass of $\mstar = 0.5~\msun$ and a core mass in the range of $1 \leq \mcore/\msun \leq 10$, the HT17 models only include Stage~0 protostars.  While the D08 models, which are based on models presented by \citet{crapsi2008:models}, do not explicitly specify \mstar\ as a model parameter, they feature the same range of core masses, the same protostellar effective temperature, and nearly the same range of protostellar luminosities as the HT17 models, thus it is highly likely that the D08 models are also dominated by Stage~0 protostars.

To explicitly test if the correlation between \fseventy\ and \lint\ holds for both Stage 0 and Stage I protostars, we repeat the above analysis separately for the models featuring Stage~0 protostars and the models featuring Stage~I protostars, with Stage~0 models defined to be those with $\left( \mstar + \mdisc \right) / \left( \mstar + \mdisc + \mcore \right) < 0.5$ and Stage~I models defined to be those with values greater than or equal to 0.5.  The results of these separate fits are tabulated in the last two columns of Table \ref{tab_f70}.  With $\sigma_{\delta_{70}}$ values of 0.249 for the Stage~0 models and 0.267 for the Stage~I models, using these fits would lead to a 1$\sigma$ scatter between estimated and intrinsic luminosities of a factor of $10^{0.249} = 1.77$ for Stage~0 protostars and a factor of $10^{0.267} = 1.84$ for Stage~I protostars.  These results are marginally better than the 1$\sigma$ scatter of a factor of $10^{0.302} = 2.00$ for the combined fit to all protostars.  However, given that these fits are most likely to be used to estimate the luminosities of protostars with limited photometric measurements that are thus difficult or impossible to separate into Class~0 and Class~I protostars, and further given the uncertainty in the assumed one-to-one correspondence between observed Class and physical Stage, it is the combined fit to all (Stage~0~+~Stage~I) protostars that will likely be the most useful for future studies of the luminosities of protostars.

\subsection{Examining correlations between flux and luminosity at all wavelengths}\label{sec_results_allwv}

Both D08 and HT17 examined correlations between $F_{\nu}$ and \lint\ at only a small number of mid-infrared wavelengths.  Specifically, D08 examined correlations at six mid-infrared wavelengths corresponding to {\it Spitzer Space Telescope} photometric bands \citep[3.6, 4.5, 5.8, 8.0, 24, and 70~\um;][]{werner2004:spitzer,fazio2004:irac,rieke2004:mips}, and HT17 examined correlations at 24 and 70 \um\ along with seven mid-infrared wavelengths corresponding to {\it Stratospheric Observatory for Infrared Astronomy (SOFIA)} photometric bands \citep[19.7, 24.2, 25.3, 31.5, 33.6, 34.8, and 37.1~\um;][]{adams2010:forcast,herter2012:forcast}.  Both studies found that the correlation featured the lowest scatter at 70~\um\ and an increasing scatter as the wavelength decreased, indicating that, of the wavelengths examined, \fseventy\ produced the most accurate estimate of the internal luminosity of a protostar.

\begin{figure*}
    \resizebox{5.0in}{!}{\includegraphics{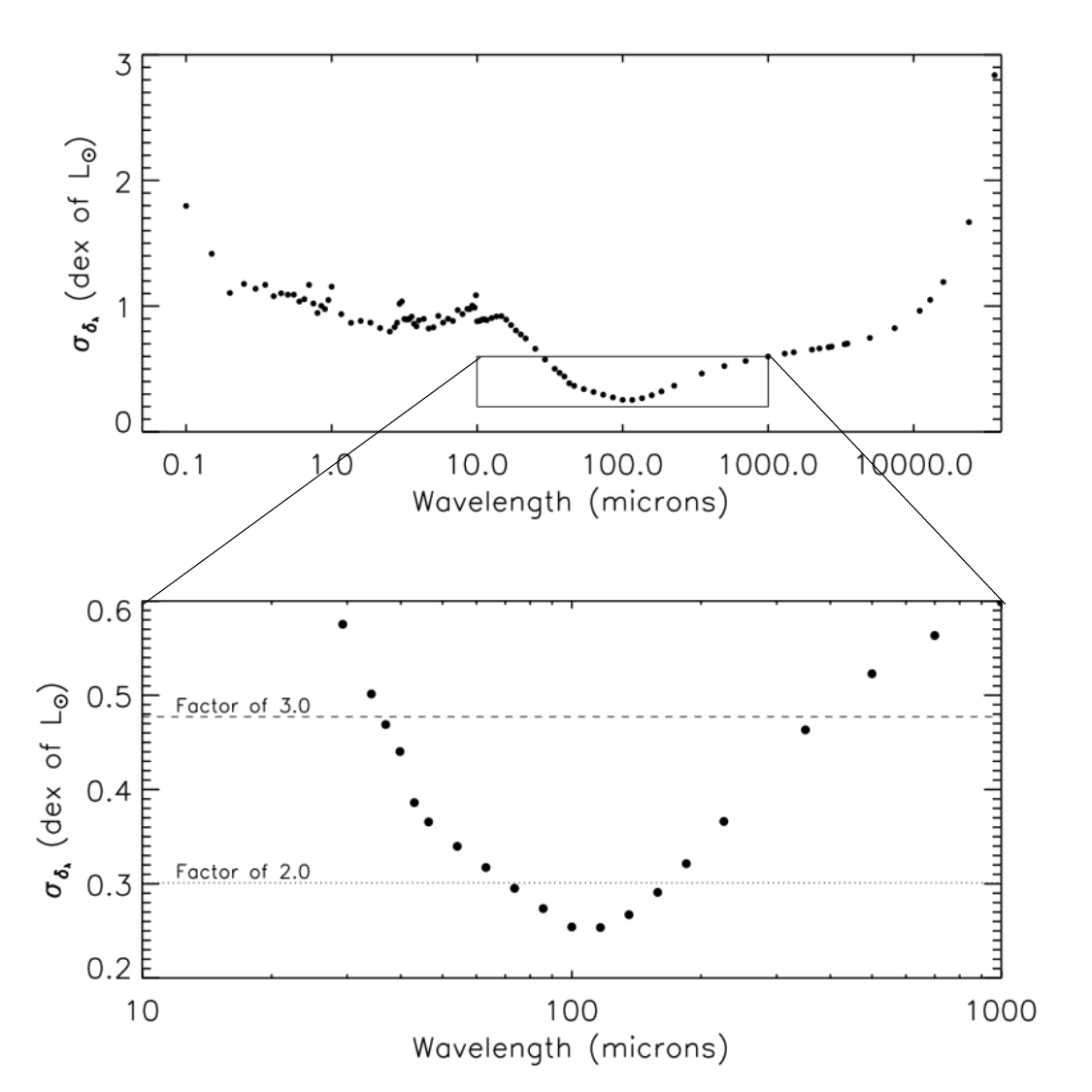}}
    \vspace{0.0in}
    \caption{\sdl~vs.~$\lambda$ for the 100 wavelengths in the radiative transfer wavelength grid, where \sdl\ is the standard deviation of $\delta_{\lambda}$, which in turn is defined according to Equation \ref{eq_delta}.  The top panel shows the full wavelength range whereas the bottom panel zooms in on the wavelength range of $10-1000$~\um.  The thin solid rectangle in the top panel shows the exact region plotted in the bottom panel.  The dotted horizontal line in the bottom panel marks the value of $\sdl = \log_{10}(2) = 0.301$ and the dashed horizontal line marks $\sdl = \log_{10}(3) = 0.477$.}
\label{fig_allwv}
\end{figure*}

The radiative transfer models presented by \citet{dunham2012:evolmodels} and used here perform the radiative transfer calculations at a grid of 100 wavelengths ranging between $0.1-36000$~\um\ (1000~\AA~--~3.6~cm).  To examine if 70~\um\ is truly the best wavelength to use, we repeat the analysis described above in \S \ref{sec_results_f70} for all 100 wavelengths in the radiative transfer models.  Tables \ref{tab_all100wv}~--~\ref{tab_all100wv_2} tabulate, for each wavelength, the best-fit values of $c_0$ and $c_1$, along with the corresponding value of \sdl, where \sdl\ is the standard deviation of $\delta_{\lambda}$, the fit deviation at each wavelength, defined following Equation \ref{eq_delta}. Additionally, Figure \ref{fig_allwv} plots \sdl\ vs.~$\lambda$ for all 100 wavelengths, with the bottom panel zooming in to show only wavelengths in the range of $10-1000$~\um.

Inspection of Tables \ref{tab_all100wv}~--~\ref{tab_all100wv_2} and Figure \ref{fig_allwv} reveal that $\sdl < \log_{10}(3)$ ($\sdl < 0.477$) for approximately $40 < \lambda/\mu m < 350$, indicating that the observed value of \fnu\ of a protostar at any single wavelength in this range can be used to estimate \lint\ for that protostar to within a 1$\sigma$ uncertainty of a factor of 3.  If we instead require that $\sdl < \log_{10}(2)$ ($\sdl < 0.301$), indicating that the observed \fnu\ of a protostar can be used to estimate \lint\ to within a 1$\sigma$ uncertainty of a factor of 2, the approximate wavelength range is $70 < \lambda/\mu m < 160$.  These findings match the general expectation that shorter wavelengths are more sensitive to geometry and viewing angle and longer wavelengths are more sensitive to the total dust mass.

\begin{figure*}
    \resizebox{6.0in}{!}{\includegraphics{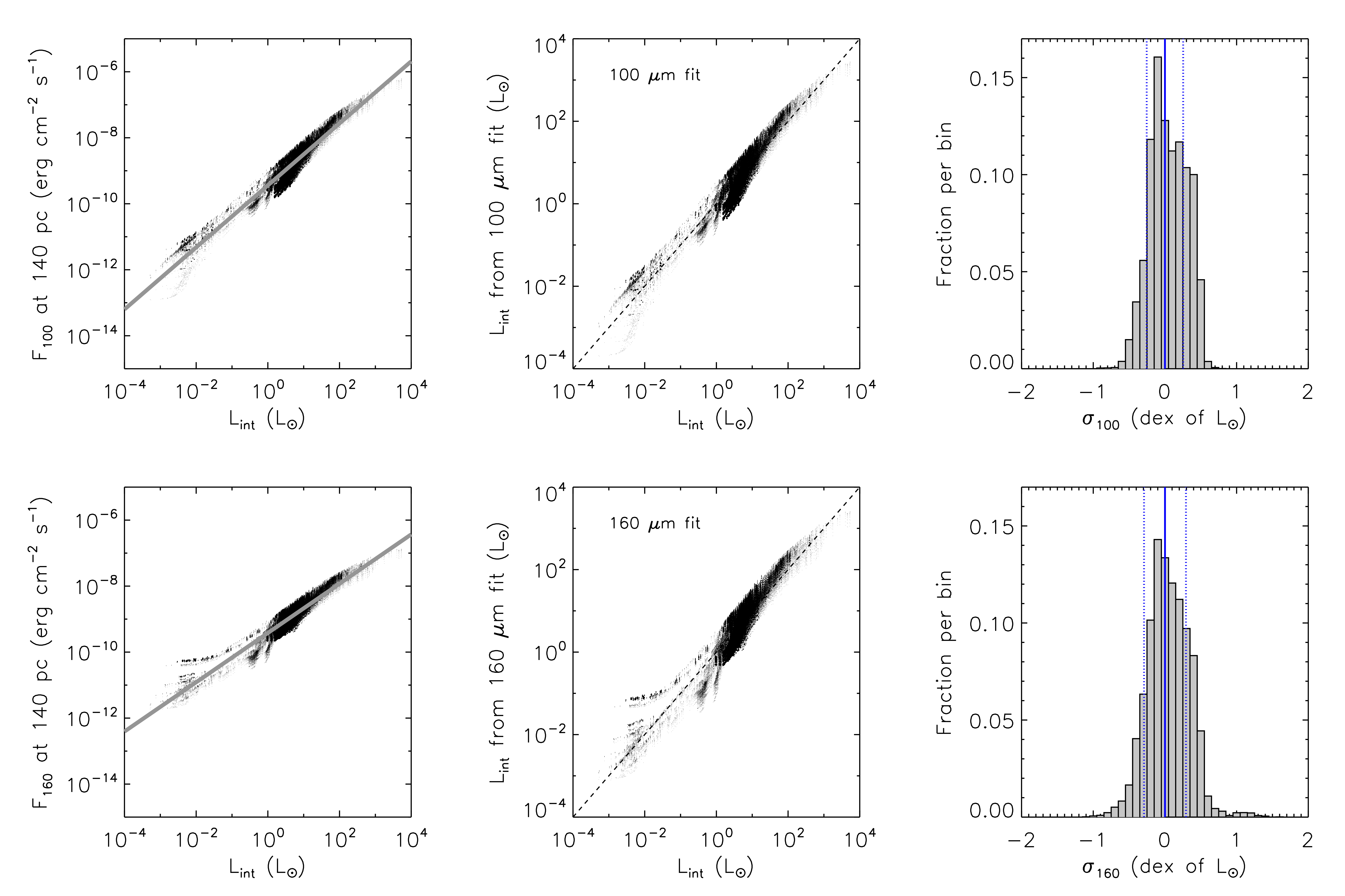}}
    \vspace{0.0in}
    \caption{\underline{Top Left:} \fhundred\ vs.~\lint\ for the 89,910 SEDs generated from our radiative transfer models.  The symbol size for each model is proportional to the value of $w_{\rm total}$ for that model, where $w_{\rm total}$ is calculated as described in \S \ref{sec_results_weights}.  The thick gray line shows the linear least-squares fit in log-log space from this current study.  \underline{Top Center:}  \lint\ estimated from the best-fit to Equation \ref{eq_fit_f70}, except using \fhundred\ instead of \fseventy, plotted vs.~the intrinsic \lint\ in the model. The symbol size for each point is proportional to the value of $w_{\rm total}$ for that model, where $w_{\rm total}$ is calculated as described in \S \ref{sec_results_weights}.  The dashed line shows the one-to-one line.  \underline{Top Right:} Weighted histogram of $\delta$ values for the \fhundred~vs.~\lint\ fit in this work, where $\delta$ is calculated as defined in Equation \ref{eq_delta} and each individual value of $\delta$ is weighted in the histogram by the $w_{\rm total}$ associated with the model that produced that $\delta$.  The solid blue vertical line indicates the mean value, and the dotted blue vertical lines indicate $\pm~1\sigma_{\rm \delta}$ from the mean.  \underline{Bottom panels:} Same as the top panels, except for 160~\um\ rather than 100~\um.}
\label{fig_f100160all}
\end{figure*}

While $\lambda = 70$~\um\ is in the range of the best possible wavelengths to use, wavelengths closer to 100~\um\ have slightly lower values of \sdl.  Given that the Photodetector Array Camera and Spectrometer \cite[PACS;][]{poglitsch2010:pacs} instrument aboard the {\it Herschel Space Observatory} obtained copious photometry of star-forming regions at 70, 100, and 160 \um\ \citep[e.g.,][]{andre2010:gb,molinari2010:higal,ragan2012:epos,launhardt2013:epos,furlan2016:hops,sadavoy2018:globules}, Figure \ref{fig_f100160all} shows linear least-squares fits in log-log space to \fnu~vs.~\lint\ at precisely 100 and 160 \um, where linear interpolation to \fnu~vs.~$\lambda$ in log-log space is used to calculate \fhundred\ and \fhsixty\ for each model SED.  The results of these fits, which are tabulated in Table \ref{tab_herschel}, demonstrate that the best {\it Herschel} wavelength to use is 100~\um, where $\sdl = 0.254$, indicating that the luminosity of a protostar can be estimated from its flux at this wavelength to within a 1$\sigma$ uncertainty of a factor of $10^{0.254}~=~1.8$.

\begin{table*}
\begin{center}
\caption{Results of linear least-squares fits to $\log{f_{\lambda}}$ vs.~$\log{\lint}$ at 70, 100, and 160~\um.}
\label{tab_herschel}
\begin{tabular}{lccc}
\hline \hline
Wavelength & $c_0 \pm \Delta~c_0$ & $c_1 \pm \Delta~c_1$ & $\sigma_{\delta}$ \\
\hline
70~\um     & $-9.541~\pm~0.001$   & $1.050~\pm~0.001$    & 0.302     \\
100~\um    & $-9.443~\pm~0.001$   & $0.940~\pm~0.001$    & 0.254     \\
160~\um    & $-9.420~\pm~0.001$   & $0.747~\pm~0.001$    & 0.293     \\
\hline 
\end{tabular}
\end{center}
\end{table*}

\subsection{Estimating Luminosity with JWST Observations}\label{sec_results_jwst}

Two decades ago, the (at the time) revolutionary increase in mid-infrared imaging sensitivity provided by the launch of the {\it Spitzer Space Telescope} enabled rapid, wide-field imaging of nearby star-forming regions, resolving individual young stellar objects and leading to a nearly complete census of star formation within 500~pc of the Sun \citep[e.g.,][]{evans2003:c2d,evans2009:c2d,rebull2010:taurus,megeath2012:orion,dunham2015:gb}.  Indeed, it was the need to quickly analyze protostars detected by the {\it Spitzer Space Telescope} that motivated the original work by \citet{dunham2008:lowlum} to identify correlations between the internal luminosity of a protostar and its flux at various wavelengths.

Currently, the {\it James Webb Space Telescope} (JWST) offers a similar revolutionary increase in near and mid-infrared imaging sensitivity and resolution, enabling rapid, wide-field imaging of distant star-forming regions that are still capable of resolving individual young stellar objects \citep[see, e.g.,][for recent examples]{izumi2024:jwst_survey,crowe2025:jwst_survey}.  However, with imaging filters at wavelengths of $0.6-5.0$~\um\ provided by the JWST Near Infrared Camera \citep[NIRCam][]{rieke2005:nircam,rieke2023:nircam} and $4.9-27.9$~\um\ provided by the JWST Mid-Infrared Instrument (MIRI), the protostellar fluxes covered by JWST do not correlate as well with the corresponding protostellar luminosities as they do for longer wavelength fluxes (see Figure \ref{fig_allwv}).

Indeed, as seen in Table \ref{tab_all100wv}--\ref{tab_all100wv_2} and Figure \ref{fig_allwv}, $\sdl$ ranges from approximately unity at the shortest JWST wavelengths to $\sdl \sim 0.6$ at the longest JWST wavelength.  Thus, in the worst-case scenario, using the observed flux in a single filter at the shortest wavelengths observed by JWST to estimate the internal luminosity of a protostar will result in a 1$\sigma$ uncertainty of approximately a factor of $10^1=10$.  Even in the best-case scenario, where the observed flux in a single filter at the longest wavelengths observed by JWST is used, the estimated internal luminosity will still be uncertain by a factor of $\sim$$10^{0.6} = 4.0$.  Observed protostellar fluxes at a single JWST wavelength produce less reliable estimates of the internal luminosities of the corresponding protostars than fluxes at longer wavelengths.  These results are consistent with HT17, who found similar results out to 37.1 \um, the longest wavelength observed by SOFIA.

Even if single wavelength protostellar fluxes in the near and mid-infrared do not exhibit tight correlations with the internal luminosity of the protostar, the use of multiple flux measurements over these wavelength ranges can still be used to obtain reliable internal luminosity estimates.  This point was first explored by \citet{kryukova2012:luminosities}, who noted that the conversion factor between mid-infrared luminosity (the integrated luminosity calculated over a narrow wavelength range of $1.2 - 24$~\um) and total luminosity depends on the infrared spectral index $\alpha$ (the slope of $\log(\lambda f_{\lambda})$ vs. $\log(\lambda)$ over the same wavelength range), since a more steeply rising SED will require a larger conversion factor between mid-infrared and total luminosity.  Indeed, they derived a quantitative relationship that allowed them to estimate the total protostellar luminosity from calculated mid-infrared luminosity and spectral index.  Defining $\sigma_{\delta}$ to again be the standard deviation of $\delta$, where $\delta$ is calculated according to Equation \ref{eq_delta}, \citet{kryukova2012:luminosities} calculate $\sigma_{\delta}=0.35$ for their relationship.  Thus their method, using only mid-infrared fluxes, gives a luminosity\footnote{While \citet{kryukova2012:luminosities} did not distinguish between bolometric and internal luminosities, the two generally agree for all but the lowest luminosity protostars, as it is only for such low luminosity protostars that the additional luminosity added from external heating becomes significant.} estimate with a reliability that is marginally worse than, but overall comparable to, our estimates using single far-infrared fluxes at $70-160$~\um.

HT17 further explored the use of shorter wavelengths by considering the mid-infrared wavelength range of $19.7-37.1$~\um\ observed by SOFIA.  They explored a simpler method, showing that a linear least squares fit between the internal luminosity of a protostar and its fluxes at \underline{\it two} SOFIA wavelengths produces relatively reliable estimates of \lint.  Specifically, they performed fits of the form\footnote{It is worth explicitly noting that these fits now treat the fluxes as independent variables, unlike Eq.\ref{eq_fit_f70}, where \lint\ is treated as the independent variable.  This change was made by HT17 for convenience of fitting using standard linear least squares techniques.}

\begin{equation}\label{eq_fit_2filter}
    \log (\lint) = c_0 + c_1 \log (F_{\nu,\lambda_1}) + c_2 \log (F_{\nu,\lambda_2}) \qquad ,
\end{equation}

\noindent with \lint\ in units of \lsun\ and all fluxes in units of erg~s$^{-1}$~cm$^{-2}$, normalized to their value at an assumed distance of 140~pc.  HT17 found that the combination of a shorter SOFIA wavelength for $\lambda_1$ (between 19.7--25.3~\um) and a longer SOFIA wavelength for $\lambda_2$ (between 31.5--37.1~\um) gave internal luminosity estimates with uncertainties comparable to those obtained using 70~\um\ alone.

{\color{black}
\subsubsection{Using Two JWST Filters}
}

Although JWST observes at shorter wavelengths than SOFIA ($0.705-25.5$~\um\ for JWST compared to $19.7-37.1$~\um\ for SOFIA), we explored applying the HT17 method to JWST observations by performing a linear least-squares fit to Equation~\ref{eq_fit_2filter} for every unique pair of wide-band JWST filters.  For NIRCam, this consists of all eight wide-band filters listed as general purpose filters in the JWST User Documentation\footnote{Available at https://jwst-docs.stsci.edu/jwst-near-infrared-camera/nircam-instrumentation/nircam-filters}, whereas for MIRI, this consists of all nine broadband imaging filters listed in the JWST User Documentation\footnote{Available at https://jwst-docs.stsci.edu/jwst-mid-infrared-instrument/miri-instrumentation/miri-filters-and-dispersers}.  Table \ref{tab_jwst_filters} lists the instrument, name, and wavelength of all 17 JWST filters considered in this work.

\begin{table}
\begin{center}
\caption{{\it JWST} filters considered in this study}
\label{tab_jwst_filters}
\begin{tabular}{lcc}
\hline \hline
           &        & $\lambda_{\rm pivot}$$^{\rm a}$ \\
Instrument & Filter & (\um)        \\
\hline
NIRCam & F070W  & 0.705 \\
NIRCam & F090W  & 0.902 \\
NIRCam & F115W  & 1.154 \\
NIRCam & F150W  & 1.501 \\
NIRCam & F200W  & 1.988 \\
NIRCam & F277W  & 2.776 \\
NIRCam & F356W  & 3.565 \\
NIRCam & F444W  & 4.402 \\
MIRI   & F560W  & 5.6   \\
MIRI   & F770W  & 7.7   \\
MIRI   & F1000W & 10.0  \\
MIRI   & F1130W & 11.3  \\
MIRI   & F1280W & 12.8  \\
MIRI   & F1500W & 15.0  \\
MIRI   & F1800W & 18.0  \\
MIRI   & F2100W & 21.0  \\
MIRI   & F2550W & 25.5  \\
\hline 
\end{tabular}
\end{center}
$^{\rm a}$The pivot wavelength, $\lambda_{\rm pivot}$, is defined in \citet{tokunaga2005:pivotlam} and is the ``central'' wavelength definition used for each filter in the JWST User Documentation.
\end{table}

\begin{figure*}
    \resizebox{5.0in}{!}{\includegraphics{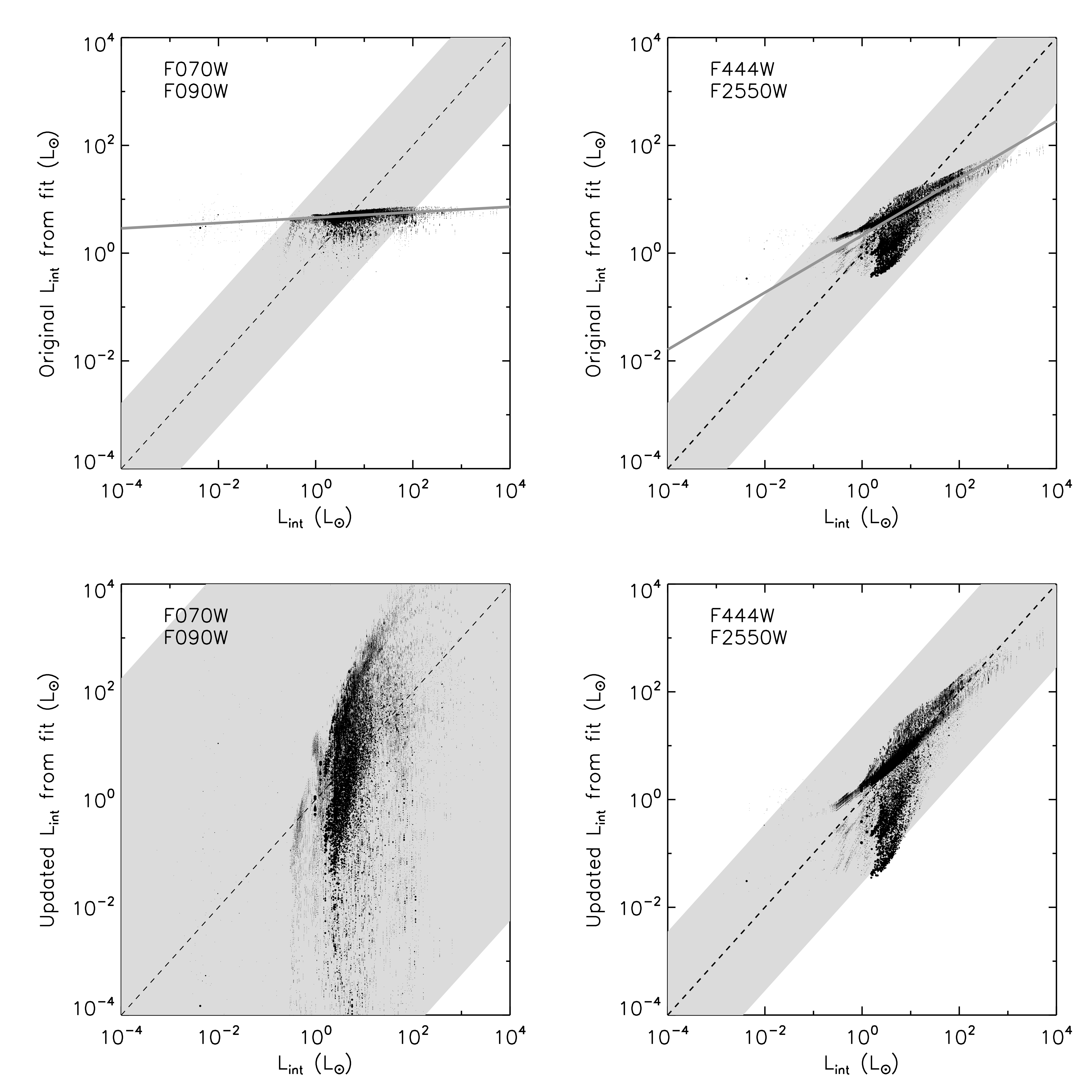}}
    \vspace{0.0in}
    \caption{\color{black}\underline{Top Left:} \lintfit\ determined from Eq.~\ref{eq_fit_2filter} using the F070W and F090W filters plotted vs.~\lintmodel, for the 89,910 model SEDs generated from our radiative transfer models.  The symbol size for each point is proportional to the value of $w_{\rm total}$ for that model, where $w_{\rm total}$ is calculated as described in \S \ref{sec_results_weights}.  The dashed line shows the one-to-one line, the solid gray line shows the best-fit linear relationship to \lintfit~vs.~\lintmodel, and the shaded gray area shows the $\pm3\sdl$ range around the one-to-one line.  \underline{Top Right:} Same as the top left panel, except for the F444W and F2550W filters.  \underline{Bottom Left:} Same as the top left panel, except after updating the fit to \lintfitc\ as described in the text.  The shaded gray area now shows the $\pm3\sdlu$ range around the one-to-one line, and the best-fit linear relationship to \lintfitc~vs.~\lintmodel\ is not shown because it sits directly on top of the one-to-one line.  \underline{Bottom Right:} Same as the bottom left panel, except for the F444W and F2550W filters.}
\label{fig_jwst_2filter}
\end{figure*}

Tables \ref{tab_jwst_2_all-a} -- \ref{tab_jwst_2_all-c} in Appendix \ref{sec_app_fulltables} report the best-fit values (and their uncertainties) for $c_0$, $c_1$, and $c_2$ for all 136 unique combinations of two JWST filters, along with the resulting value of \sdl\ for each fit.  {\color{black}In a multi-dimensional linear fit such as that represented by Equation \ref{eq_fit_2filter}, while the least-squares method ensures that the resulting fit minimizes the residuals between the intrinsic and fit values of $\log(\lint)$, there is no guarantee the linear fit between these quantities that results from this minimization will have a slope of unity, in which case the fit could either systematically under-predict or over-predict the actual \lint\ of the object in various luminosity ranges.  To illustrate this point, the top panels of Figure \ref{fig_jwst_2filter} plot, for two different filter combinations (F070W and F090W, and F444W and F2550W), the estimated \lint\ from the fit (hereafter referred to as \lintfit) vs.~the intrinsic \lint\ for each model (hereafter referred to as \lintmodel).  With $\sdl=0.408$ in each case, as tabulated in Tables \ref{tab_jwst_2_all-a} -- \ref{tab_jwst_2_all-c}, the two filter combinations give fits with identical scatter despite the former combination significantly over-predicting \lintfit\ at low intrinsic \lintmodel\ and significantly under-predicting \lintfit\ at high intrinsic \lintmodel.
}

{\color{black}To quantify and correct for this effect, after fitting to Eq.~\ref{eq_fit_2filter} for each possible combination of two JWST filters to determine $c_0$, $c_1$, and $c_2$, we apply our fit parameters to calculate \lintfit\ from the model fluxes for each of the 89,910 models.  We then perform the following linear least-squares fit between \lintfit\ and the intrinsic \lintmodel\ for each model:

\begin{equation}\label{eq_fit_lvl}
\log \left( \lintfit \right) = m_{\rm LvL} \log \left( \lintmodel \right) + b_{\rm LvL} \qquad ,
\end{equation}
where the ${\rm LvL}$ subscripts stand for ``luminosity vs.~luminosity.''  We then use the best-fit $m_{\rm LvL}$ and $b_{\rm LvL}$ to correct each \lintfit:

\begin{equation}\label{eq_fit_correct}
\log \left( \lintfitc \right) = \frac{\log \left( \lintfit \right) - b_{\rm LvL}}{m_{\rm LvL}} \qquad .
\end{equation}
Finally, we calculate \sdlu\ by comparing the corrected \lintfitc\ to \lintmodel.  The last three columns of \ref{tab_jwst_2_all-a} -- \ref{tab_jwst_2_all-c} in Appendix \ref{sec_app_fulltables} report the best-fit values of $m_{\rm LvL}$ and $b_{\rm LvL}$, along with the resulting \sdlu.

The bottom panels of Figure~\ref{fig_jwst_2filter} plot the updated \lintfitc\ vs.~\lintmodel\ for the two filter combinations with identical \sdl\ values considered previously (F070W and F090W, and F444W and F2550W).  In both cases the \lintfitc\ and \lintmodel\ now exhibit perfect one-to-one correlations\footnote{\color{black}Even though a ``by-eye fit'' to the bottom left panel of Figure~\ref{fig_jwst_2filter} might suggest a very steep slope, the best-fit line obtained via linear least squares truly does have a slope identical to unity and an intercept identical to zero.}, but with \sdlu\ values that have increased compared to the original \sdl\ values (in the former case by a very large amount).  These increased values of \sdlu\ (indicating increased scatter between the intrinsic and fit values of \lint) are to be expected given that the original fits before correction were those that minimized the scatter.  This increased scatter is a necessary consequence of ensuring a one-to-one correlation between intrinsic and fit values of \lint\ across the full protostellar luminosity range.

Table \ref{tab_jwst_2_best} reports the same information as Tables \ref{tab_jwst_2_all-a} -- \ref{tab_jwst_2_all-c} in Appendix \ref{sec_app_fulltables} for the ten best combinations of two JWST filters (those models with the ten lowest values of \sdlu), and the left panel of Figure \ref{fig_jwst_multiple} plots \lintfitc\ vs.~\lintmodel\ for the best combination of two JWST filters (F356W and F2550W).  With \sdlu\ values ranging between 0.493 and 0.544, corresonding to 1$\sigma$ uncertainties in \lintfitc\ of factors of $10^{0.493} (3.1)$ -- $10^{0.544} (3.5)$, even the best combinations of two JWST filters perform somewhat worse than single far-infrared wavelengths (for $70-160$~\um, $0.254 \leq \sdl \leq 0.302$; see \S \ref{sec_results_allwv} for details).

\begin{table*}
\begin{center}
\caption{The ten best combinations of two {\it JWST} filters}
\label{tab_jwst_2_best}
\begin{tabular}{ccccccccc}
\hline \hline
$\lambda_1$ (\um) & $\lambda_2$ (\um) & $c_0 \pm \Delta~c_0$ & $c_1 \pm \Delta~c_1$ & $c_2 \pm \Delta~c_2$ & $\sigma_{\delta}$ & $m_{\rm LvL}$ & $b_{\rm LvL}$ & $\sigma_{\delta}^{\rm updated}$ \\
\hline
 3.566 & 25.5 &  3.573 $\pm$  0.013 & -0.089 $\pm$  0.002 &  0.395 $\pm$  0.002 &  0.391 &  0.54 &  0.32 &  0.493 \\
  21.0 & 25.5 &  5.336 $\pm$  0.013 & -0.973 $\pm$  0.008 &  1.482 $\pm$  0.008 &  0.446 &  0.71 &  0.20 &  0.507 \\
 7.7   & 25.5 &  3.740 $\pm$  0.013 & -0.137 $\pm$  0.002 &  0.469 $\pm$  0.002 &  0.413 &  0.56 &  0.30 &  0.511 \\
 2.776 & 25.5 &  3.832 $\pm$  0.013 & -0.041 $\pm$  0.002 &  0.372 $\pm$  0.002 &  0.403 &  0.54 &  0.33 &  0.513 \\
 4.401 & 25.5 &  3.644 $\pm$  0.014 & -0.063 $\pm$  0.002 &  0.377 $\pm$  0.002 &  0.408 &  0.53 &  0.33 &  0.517 \\
 1.154 & 25.5 &  3.432 $\pm$  0.013 & -0.016 $\pm$  0.001 &  0.303 $\pm$  0.001 &  0.362 &  0.45 &  0.41 &  0.526 \\
 0.704 & 25.5 &  3.401 $\pm$  0.013 &  0.009 $\pm$  0.001 &  0.274 $\pm$  0.001 &  0.354 &  0.43 &  0.43 &  0.529 \\
 5.6   & 25.5 &  3.649 $\pm$  0.014 & -0.107 $\pm$  0.002 &  0.425 $\pm$  0.002 &  0.409 &  0.52 &  0.35 &  0.532 \\
18.0   & 25.5 &  5.043 $\pm$  0.013 & -0.431 $\pm$  0.004 &  0.906 $\pm$  0.005 &  0.463 &  0.68 &  0.23 &  0.537 \\
15.0   & 25.5 &  4.826 $\pm$  0.013 & -0.310 $\pm$  0.003 &  0.763 $\pm$  0.004 &  0.465 &  0.67 &  0.24 &  0.544 \\
\hline 
\end{tabular}
\end{center}
\end{table*}

\begin{figure*}
    \resizebox{5.0in}{!}{\includegraphics{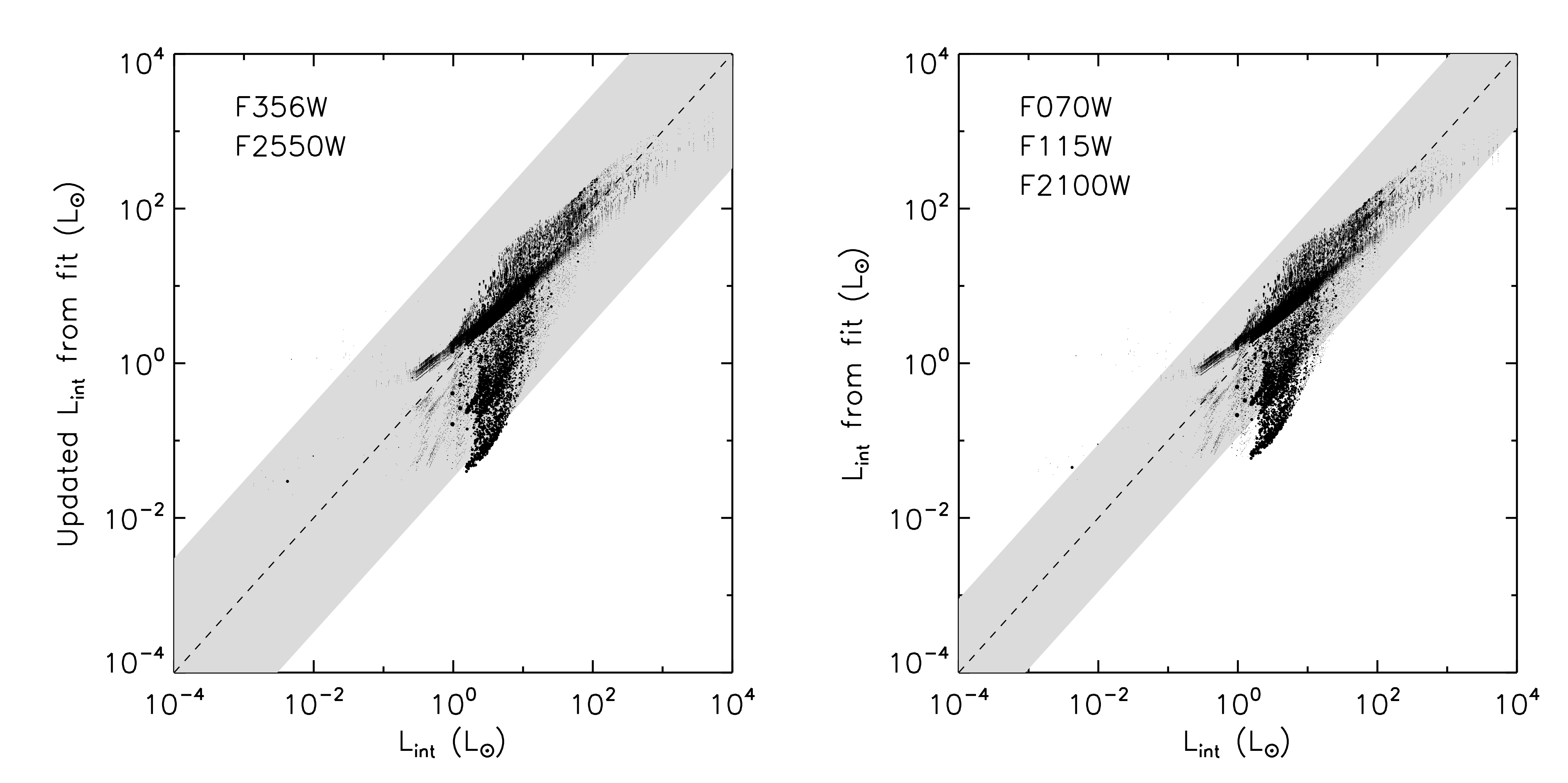}}
    \vspace{0.0in}
    \caption{\color{black}\underline{Left:} \lintfitc\ estimated from our fit to Eq.~ \ref{eq_fit_2filter}, and then updated using Eq.~\ref{eq_fit_correct}, plotted vs.~the intrinsic \lintmodel, using the combination of two JWST filters that gives the lowest value of \sdlu\ (F356W and F2550W) for the 89,910 model SEDs generated from our radiative transfer models.  The symbol size for each point is proportional to the value of $w_{\rm total}$ for that model, where $w_{\rm total}$ is calculated as described in \S \ref{sec_results_weights}.  The dashed line shows the one-to-one line and the shaded gray area shows the $\pm3\sdlu$ range around the one-to-one line.  \underline{Right:} Same as the left panel, except for the fit to Equation \ref{eq_fit_3filter} (and then again updated using Eq.~\ref{eq_fit_correct}) using the combination of three JWST filters that gives the lowest value of \sdlu\ (F070W, F115W, and F2100W).}
\label{fig_jwst_multiple}
\end{figure*}

We recommend the following procedure when using these correlations to estimate protostellar luminosities from fluxes observed in two JWST filters: first calculate \lintfit\ using Eq.~\ref{eq_fit_2filter}, and then correct this to \lintfitc\ using Eq.~\ref{eq_fit_correct}.  Even though this second step leads to increased scatter, it ensures that the estimated luminosities are correct {\it on average} across the full luminosity range, and it has the advantage of allowing \sdlu\ to be treated as a symmetrical $1\sigma$ error bar for each value of $\log \left( \lintfitc \right)$.

}

{\color{black}

\subsubsection{Using Three JWST Filters}
}

Given that large photometric surverys of star-forming regions are often conducted in multiple filters, we explore the possibility of {\color{black}improving the ability to estimate \lint\ from observed infrared fluxes by considering \underline{\it three} JWST filters rather than two filters.}  Specifically, we perform a linear least-squares fit to the function

\begin{equation}\label{eq_fit_3filter}
    \log (\lint) = c_0 + c_1 \log (F_{\nu,\lambda_1}) + c_2 \log (F_{\nu,\lambda_2}) + c_3 \log (F_{\nu,\lambda_3}) \qquad ,
\end{equation}

\noindent with \lint\ in units of \lsun\ and all fluxes in units of erg~s$^{-1}$~cm$^{-2}$, normalized to their value at an assumed distance of 140~pc.  {\color{black}Following our procedure in the previous section, we then use Eq.~\ref{eq_fit_lvl} to obtain $m_{\rm LvL}$ and $b_{\rm LvL}$, followed by Eq.~\ref{eq_fit_correct} to obtain \lintfitc\ for each model.
}

Tables \ref{tab_jwst_3_all-a} -- \ref{tab_jwst_3_all-k} in Appendix \ref{sec_app_fulltables} report the best-fit values (and their uncertainties) for $c_0$, $c_1$, $c_2$, and $c_3$ for all 680 unique combinations of three JWST filters, {\color{black}along with 
\sdl, $m_{\rm LvL}$, $b_{\rm LvL}$, and \sdlu.}
Table \ref{tab_jwst_3_best} reports the same information for the ten unique combinations of three JWST filters with the lowest values of {\color{black}\sdlu}, and the right panel of Figure \ref{fig_jwst_multiple} plots, for the combination of three filters with the lowest {\color{black}\sdlu} (F070W, F115W, and F2100W), the estimated {\color{black}\lintfitc} vs.~the intrinsic {\color{black}\lintmodel} for each model. 

\begin{table*}
\begin{center}
\caption{The best combinations of three {\it JWST} filters}
\label{tab_jwst_3_best}
\begin{tabular}{ccccccccccc}
\hline \hline
$\lambda_1$ (\um) & $\lambda_2$ (\um) & $\lambda_3$ (\um) & $c_0 \pm \Delta~c_0$ & $c_1 \pm \Delta~c_1$ & $c_2 \pm \Delta~c_2$ & $c_3 \pm \Delta~c_3$ & $\sigma_{\delta}$ & $m_{\rm LvL}$ & $b_{\rm LvL}$ & $\sigma_{\delta}^{\rm updated}$\\
\hline
 \hline
 0.704 &  1.154 &   21.0 &  1.776 $\pm$  0.405 &  0.748 $\pm$  0.009 & -1.074 $\pm$  0.003 &  0.405 $\pm$  0.004 &  0.237 &  0.59 &  0.29 &   0.318 \\
 0.704 &  1.154 &   25.5 &  1.684 $\pm$  0.204 &  0.432 $\pm$  0.011 & -0.546 $\pm$  0.003 &  0.204 $\pm$  0.004 &  0.286 &  0.33 &  0.47 &   0.322 \\
 0.704 &  0.903 &   21.0 &  1.783 $\pm$  0.293 &  1.180 $\pm$  0.010 & -1.390 $\pm$  0.006 &  0.293 $\pm$  0.007 &  0.266 &  0.47 &  0.37 &   0.342 \\
 0.704 &  1.501 &   25.5 &  2.058 $\pm$  0.239 &  0.255 $\pm$  0.011 & -0.360 $\pm$  0.002 &  0.239 $\pm$  0.003 &  0.297 &  0.35 &  0.46 &   0.343 \\
 0.704 &  1.154 &   18.0 &  1.710 $\pm$  0.593 &  1.037 $\pm$  0.010 & -1.582 $\pm$  0.003 &  0.593 $\pm$  0.005 &  0.283 &  0.81 &  0.14 &   0.343 \\
 0.704 &  1.501 &   21.0 &  2.072 $\pm$  0.463 &  0.455 $\pm$  0.010 & -0.802 $\pm$  0.002 &  0.463 $\pm$  0.003 &  0.270 &  0.62 &  0.27 &   0.354 \\
 0.704 &  0.903 &   18.0 &  1.884 $\pm$  0.446 &  1.657 $\pm$  0.010 & -2.030 $\pm$  0.006 &  0.446 $\pm$  0.007 &  0.276 &  0.67 &  0.24 &   0.361 \\
 0.704 &  1.988 &   21.0 &  2.043 $\pm$  0.476 &  0.263 $\pm$  0.010 & -0.618 $\pm$  0.002 &  0.476 $\pm$  0.003 &  0.271 &  0.57 &  0.31 &   0.367 \\
 0.903 &  1.154 &   21.0 &  1.938 $\pm$  0.382 &  1.310 $\pm$  0.010 & -1.585 $\pm$  0.007 &  0.382 $\pm$  0.008 &  0.281 &  0.53 &  0.34 &   0.376 \\
 0.704 &  1.988 &   25.5 &  2.054 $\pm$  0.205 &  0.134 $\pm$  0.012 & -0.201 $\pm$  0.002 &  0.205 $\pm$  0.003 &  0.317 &  0.29 &  0.50 &   0.381 \\
\hline 
\end{tabular}
\end{center}
\end{table*}

Estimating the internal luminosity of a protostar using the measured protostellar fluxes in three different JWST filters can produce a significantly more reliable estimate than using the measured fluxes in two filters.  Indeed, the ten best combinations of three JWST filters result give {\color{black}\sdlu} values ranging between {\color{black}$0.318-0.381$}, all of which are {\color{black}significantly} better than the lowest value of {\color{black}$\sdl=0.493$} when two filters are used.  With {\color{black}$\sdlu=0.318$}, the best possible combination of three filters (F070W, F115W, and F2100W) can be used to obtain estimated internal luminosities with a 1$\sigma$ uncertainty of {\color{black}a factor of $10^{0.318}=2.1$.  This is only marginally worse than the 1$\sigma$ uncertainty in estimated internal luminosities of factors of $10^{0.254} (1.8) -- 10^{0.302} (2.0)$ that can be obtained from single far-infrared flux measurements (see \S \ref{sec_results_f70} and \ref{sec_results_allwv} for details), illustrating that the right combinations of three optical, near-infrared, and mid-infrared JWST fluxes can provide nearly as reliable estimates of protostellar internal luminosities as single far-infrared {\it Spitzer} and {\it Herschel}  flux measurements,} opening the door to expanded catalogs of protostellar luminosities in star-forming regions too distant for individual protostars to be detected and/or resolved by {\it Spitzer} or {\it Herschel}.

{\color{black}
\section{Limitations and Future Work}\label{sec_limitations}

\subsection{Scattering}\label{sec_limitations_scattering}
Scattering dominates the total (absorption+scattering) dust opacity over most of the wavelength range observed by the JWST, as seen in both Fig.~1 of \citet{young05:evolmodels} and Fig.~4 of \citet{dunham2010:evolmodels}.  It is thus essential to include scattering to accurately predict model fluxes at JWST wavelengths.  While we do include scattering in our radiative transfer models, due to technical limitations at the time the models were generated the scattering was assumed to be isotropic when in reality the dust grains in dense cores preferentially forward-scatter light \citep[see][for an extensive discussion of this issue and the resulting impacts on the radiative transfer models]{young05:evolmodels,dunham2010:evolmodels}.

Given that our radiative transfer models adopt \underline{\it isotropic} scattering to model \underline{\it isolated} dense cores, the model fluxes at short wavelengths are uncertain in the following ways:

\begin{enumerate}
    \item The models will back-scatter too many short-wavelength photons from the interstellar radiation field back toward the observer, causing the model fluxes at these wavelengths to be too high.
    \item The models will forward-scatter too few short-wavelength photons from the protostar forward out of the core and toward the observer, causing the model fluxes at these wavelengths to be too low.
    \item The models do not include scattered light off the ambient cloud medium in the foreground of the core.  This will cause the model fluxes to be too low compared to observed fluxes (or perhaps more accurately stated, it will cause the observed fluxes to be too high compared to the model fluxes).
\end{enumerate}

Due to these competing effects, observed protostellar fluxes at wavelengths where scattering is important could be either higher or lower than our model fluxes, with the dominant effect impossible to determine from our existing models.

In an attempt to quantify the uncertainty added by these effects, we first add random scatter to our fluxes as follows:

\begin{enumerate}
    \item At $\lambda < 1.0$~\um, we allow for up to a factor of 3 uncertainty due to scattering by multiplying each model flux by a randomly chosen factor between $1/3$ and 3.
    \item At $1.0$~\um~$\leq \lambda < 5.0$~um, we allow for up to a factor of 2 uncertainty due to scattering by multiplying each model flux by a randomly chosen factor between $1/2$ and 2.
    \item At $5.0$~\um~$\leq \lambda \leq 20.0$~um, we allow for up to a factor of 1.5 uncertainty due to scattering by multiplying each model flux by a randomly chosen factor between $1/1.5$ and 1.5.
\end{enumerate}
The factors of 3, 2, and 1.5 for the random scatter are admittedly somewhat arbitrary and somewhat large, but are chosen to conservatively assess the maximum possible uncertainty added to our results by the limited treatment of scattering in our models.  They are also chosen to maximize the scatter added at the shortest wavelengths where scattering is likely most significant.  

Repeating our analysis for combinations of three JWST filters, we find that \sdlu\ features a mean increase of 0.18 with this random scatter included in the fluxes.  If we restrict ourselves to filter combinations that originally had $\sdlu \leq 0.5$, we find that \sdlu\ features a mean increase of 0.13.  Thus these filter combinations can still be used to estimate the internal luminosities of protostars, but with uncertainties that may be 0.1--0.2~dex larger than suggested by our tabulated values of \sdlu.  Future work should be devoted to constructing and utilizing new radiative transfer models that more accurately include the effects of scattering off dust grains.

\subsection{Lack of Higher Initial Core Mass Models}\label{sec_limitations_mcore}

As noted in \S \ref{sec_results_f70}, the vast majority of our models (84,807 out of 89,910, or 94\%) feature $\mcore \leq 1.0\ \msun$, and none have initial core masses greater than 2 \msun.  In contrast, the earlier work on this topic by both D08 and HT17 featured core masses in the range of $1.0 \leq \mcore/\msun \leq 10.0$.  Our extension to lower-mass cores forming lower-mass stars both increased the scatter of the resulting $\log (\fseventy)$ vs.~$\log (\lint)$ linear least-squares fit and led to a slightly revised correlation between $\log (\fseventy)$ and $\log (\lint)$.  While we don't expect any significant changes to our results if we included cores with initial masses greater than 2 \msun, since these higher-mass core models would form high-mass stars that would be significantly down-weighted by their corresponding values of $w_{\rm IMF}$ (where $w_{\rm IMF}$ is defined in \S \ref{sec_results_weights}), future work should nevertheless be devoted to exploring the full effects of including such higher-mass models.

}

\section{Summary}\label{sec_summary}

The luminosities of protostars provide an indirect method of measuring the masses and mass accretion rates of protostars in their earliest stages of evolution, but accurate measurements of protostellar internal luminosities traditionally require assembling complete spectral energy distributions (SEDs) from the near-infrared through millimeter wavelengths.  In this paper, we used published evolutionary radiative transfer models of collapsing protostellar cores to evaluate the extent to which the internal luminosity of a protostar, \lint, can be estimated with a limited number of infrared photometric measurements.  This work represents a confirmation of, and update to, the previous work of \citet{dunham2008:lowlum} (D08) and \citet{huard2017:luminosities} (HT17) that focused on mid and far-infrared photometric measurements, as well as an extension of this work to the near-infrared wavelengths observed by the {\it James Webb Space Telescope} (JWST).  We summarize our main findings as follows: 

\begin{enumerate}
    \item We confirm previous results showing a tight correlation (in log-log space) between the internal luminosity of a protostar (\lint) and its flux at 70 microns (\fseventy).  If this correlation is used to estimate \lint\ from a measured value of \fseventy, the resulting \lint\ has a 1$\sigma$ uncertainty of a factor of 2.0 (or 0.302 dex of \lsun).
    \item Our updated correlation between \lint\ and \fseventy\ yields internal luminosity estimates that are a factor of 1.9 higher, on average, than those obtained using the correlation presented by HT17, and a factor of 3.2 higher, on average, than those obtained using the correlation presented by D08.  The primary explanation for these differences is the larger range of model parameters used in the present study.
    \item The internal luminosity estimates obtained from measured values of \fseventy\ are marginally less uncertain if separate correlations are used for Stage 0 and Stage I protostars, with the 1$\sigma$ uncertainty in the estimated \lint\ decreasing to a factor of 1.77 (0.249 dex of \lsun) for Stage 0 and 1.84 (0.267 dex of \lsun) for Stage I.  However, given the marginal nature of these improvements and the difficulty in classifying protostars in cases where limited photometric measurements are available, we suggest the combined (Stage 0 + Stage I) correlation is likely the most useful for future studies.
    \item Similar correlations between \lint\ and \fnu\ can be identified at other wavelengths than 70~\um.  We present correlations at 100 wavelengths in the range of 0.10~\um~--~36000~\um~(1000~\AA~--~3.6~cm).  Any single flux measurement in the wavelength range of 40~\um~--~350~\um\ will provide an internal luminosity estimate with a 1$\sigma$ uncertainty of a factor of 3 (0.477 dex of \lsun) or lower, and any single flux measurement in the wavelength range of 70~\um~--~160~\um\ will provide an internal luminosity estimate with a 1$\sigma$ uncertainty of a factor of 2 (0.301 dex of \lsun) or lower.  While the best wavelength to use is 100~\um, where the resulting internal luminosity estimate has a 1$\sigma$ uncertainty of a factor of 1.8 (0.254 dex of \lsun), the improvement over the widely used wavelength of 70~\um\ is quite modest.
    \item Fluxes at each of the shorter wavelengths observed by JWST (0.6~--~27.9~\um) do not correlate well with \lint, with 1$\sigma$ uncertainties in the \lint\ that would be estimated from observed fluxes ranging between factors of 10 at the shortest wavelengths to factors of 4 at the longest wavelengths.
    \item Using {\color{black}a single photometric measurement} in two different JWST filters simultaneously can improve the reliability of the resulting estimate for \lint\ {\color{black}compared to using observations in only a single JWST filter}.  The best combination of two filters ({\color{black}F356W} and F2550W) results in a 1$\sigma$ uncertainty in the resulting \lint\ estimate of a factor of {\color{black}3.1 (0.493 dex in \lsun), better than even the lowest uncertainty when using a single JWST filter.}
    \item Using {\color{black}a single photometric measurement} in three different JWST filters simultaneously can further improve the reliability of the resulting estimate for \lint.  The best combination of three filters (F070W, F115W, and F2550W) results in a 1$\sigma$ uncertainty in the resulting \lint\ estimate of a factor of {\color{black}2.1 (0.318 dex in \lsun)}.  This is {\color{black}comparable to the uncertainties of factors of $1.8-2.0$ that can be obtained from single far-infrared flux measurements.}
\end{enumerate}

The correlations between observed infrared fluxes of protostars and the underlying internal luminosities of those protostars that we present here can be used to obtain reliable estimates of large samples of protostars without the need to assemble complete spectral energy distributions across several orders of magnitude in wavelength.  The JWST correlations may be particularly useful as current and future JWST surveys of star-forming regions detect thousands of protostars in the near and mid-infrared, in distant regions where the sensitivitiy and angular resolution of ground and space-based far-infrared and submillimeter telescopes are insufficient to detect and resolve each individual protostar.  {\color{black}Future work should revisit and confirm (or update, if necessary) these correlations with models that more accurately include the effects of scattering of light off dust grains and better sample the full range of the dense core mass function.}

\section*{Acknowledgements}

MMD acknowledges support from NSF award AST-2508249.  NH is funded by the Spanish grant MCIN/AEI/10.13039/501100011033 PID2023-150468NB-I00.  AP acknowledges financial support from the UNAM-PAPIIT IN120226 grant, and the Sistema Nacional de Investigadores of SECIHTI, M\'exico.  The authors acknowledge the intellectual contributions provided by Antonio Crapsi, who first had the idea to use radiative transfer models to test empirical correlations between observed fluxes and known luminosities of protostars.  {\color{black}We also thank the referee for comments that have improved the quality of this work.}

This research has made use of NASA’s Astrophysics Data System (ADS) Abstract Service and the IDL Astronomy Library hosted by the NASA Goddard Space Flight Center. 

\section*{Data Availability}

No new data were used in the preparation of this publication.  Both the fluxes and the internal luminosities for the model protostars are taken from the evolutionary radiative transfer models published previously by \citet{dunham2012:evolmodels}, which rely on hydrodynamical simulations first published in \citet{vorobyov2005:bursts, vorobyov2006:bursts, vorobyov2010:bursts}.



\bibliographystyle{mnras}
\bibliography{dunham_citations}




\appendix

\section{Full Results Tables}\label{sec_app_fulltables}

Tables \ref{tab_all100wv}~--~\ref{tab_all100wv_2} tabulate, for each wavelength, the best-fit values of $c_0$ and $c_1$ from Equation \ref{eq_fit_f70}, along with the corresponding value of \sdl, where \sdl\ is the standard deviation of $\delta_{\lambda}$, the fit deviation at each wavelength defined following Equation \ref{eq_delta}.

Tables \ref{tab_jwst_2_all-a}~--~\ref{tab_jwst_2_all-c} report the best-fit values (and their uncertainties) for $c_0$, $c_1$, and $c_2$ in Equation \ref{eq_fit_2filter} for all 136 unique combinations of two JWST filters, along with the resulting value of \sdl\ for each fit.  Tables \ref{tab_jwst_3_all-a}~--~\ref{tab_jwst_3_all-k} report the best-fit values (and their uncertainties) for $c_0$, $c_1$, $c_2$, and $c_3$ in Equation \ref{eq_fit_3filter} for all 680 unique combinations of three JWST filters, along with the resulting value of \sdl\ for each fit.  

\begin{table*}
\begin{center}
\caption{Results of linear least-squares fits to $\log{F_{\\nu}}$ vs.~$\log{\lint}$}
\label{tab_all100wv}
\begin{tabular}{lccc}
\hline \hline
Wavelength (\um) & $c_0 \pm \Delta~c_0$ & $c_1 \pm \Delta~c_1$ & $\sigma_{\delta}$ \\
\hline
    0.10  &  $-17.034 \pm 0.003$  &  $0.308 \pm 0.004$  &  1.797  \\
    0.15  &  $-15.652 \pm 0.005$  &  $0.655 \pm 0.006$  &  1.417  \\
    0.20  &  $-14.579 \pm 0.007$  &  $1.194 \pm 0.009$  &  1.105  \\
    0.25  &  $-13.743 \pm 0.007$  &  $1.255 \pm 0.008$  &  1.177  \\
    0.30  &  $-12.977 \pm 0.007$  &  $1.321 \pm 0.009$  &  1.138  \\
    0.35  &  $-12.432 \pm 0.008$  &  $1.315 \pm 0.009$  &  1.170  \\
    0.40  &  $-11.986 \pm 0.007$  &  $1.362 \pm 0.008$  &  1.079  \\
\hline 
\end{tabular}
\end{center}
\end{table*}

\begin{table*}
\begin{center}
\caption{Continuation of Table \ref{tab_all100wv}.}
\label{tab_all100wv_2}
\begin{tabular}{lccc}
\hline \hline
Wavelength (\um) & $c_0 \pm \Delta~c_0$ & $c_1 \pm \Delta~c_1$ & $\sigma_{\delta}$ \\
\hline
   18.48  &  $-10.818 \pm 0.004$  &  $1.248 \pm 0.005$  &  0.806  \\
   19.95  &  $-10.750 \pm 0.004$  &  $1.258 \pm 0.005$  &  0.774  \\
   21.55  &  $-10.681 \pm 0.004$  &  $1.263 \pm 0.004$  &  0.742  \\
   25.12  &  $-10.518 \pm 0.004$  &  $1.257 \pm 0.004$  &  0.661  \\
   29.29  &  $-10.341 \pm 0.003$  &  $1.223 \pm 0.003$  &  0.575  \\
   34.15  &  $-10.154 \pm 0.002$  &  $1.191 \pm 0.003$  &  0.501  \\
   36.87  &  $-10.059 \pm 0.002$  &  $1.183 \pm 0.003$  &  0.469  \\
\hline 
\end{tabular}
\end{center}
\end{table*}

\begin{table*}
\begin{center}
\caption{All possible combinations of two {\it JWST} filters}
\label{tab_jwst_2_all-a}
\begin{tabular}{ccccccccc}
\hline \hline
$\lambda_1$ (\um) & $\lambda_2$ (\um) & $c_0 \pm \Delta~c_0$ & $c_1 \pm \Delta~c_1$ & $c_2 \pm \Delta~c_2$ & $\sigma_{\delta}$ & $m_{\rm LvL}$ & $b_{\rm LvL}$ & $\sigma_{\delta}^{\rm updated}$ \\
\hline
 0.704 &  0.903 &  1.024 $\pm$  0.014 &  0.385 $\pm$  0.008 & -0.357 $\pm$  0.009 &  0.408 &  0.05 &  0.66 &  2.085 \\
 0.704 &  1.154 &  1.019 $\pm$  0.014 &  0.185 $\pm$  0.004 & -0.154 $\pm$  0.005 &  0.405 &  0.04 &  0.67 &  2.220 \\
 0.704 &  1.501 &  1.367 $\pm$  0.015 &  0.082 $\pm$  0.003 & -0.011 $\pm$  0.004 &  0.418 &  0.04 &  0.67 &  2.200 \\
 0.704 &  1.988 &  1.548 $\pm$  0.015 &  0.045 $\pm$  0.002 &  0.046 $\pm$  0.003 &  0.416 &  0.05 &  0.67 &  1.931 \\
 0.704 &  2.776 &  1.764 $\pm$  0.015 &  0.035 $\pm$  0.002 &  0.078 $\pm$  0.002 &  0.418 &  0.07 &  0.66 &  1.582 \\
 0.704 &  3.566 &  1.786 $\pm$  0.015 &  0.030 $\pm$  0.002 &  0.084 $\pm$  0.002 &  0.410 &  0.08 &  0.66 &  1.400 \\
 0.704 &  4.401 &  1.888 $\pm$  0.015 &  0.029 $\pm$  0.001 &  0.094 $\pm$  0.002 &  0.408 &  0.09 &  0.65 &  1.286 \\
\hline 
\end{tabular}
\end{center}
\end{table*}
\begin{table*}
\begin{center}
\caption{Continuation of Table \ref{tab_jwst_2_all-a}}
\label{tab_jwst_2_all-b}
\begin{tabular}{ccccccccc}
\hline \hline
$\lambda_1$ (\um) & $\lambda_2$ (\um) & $c_0 \pm \Delta~c_0$ & $c_1 \pm \Delta~c_1$ & $c_2 \pm \Delta~c_2$ & $\sigma_{\delta}$ & $m_{\rm LvL}$ & $b_{\rm LvL}$ & $\sigma_{\delta}^{\rm updated}$ \\
\hline
 1.988 &  7.7   &  2.297 $\pm$  0.014 &  0.071 $\pm$  0.003 &  0.093 $\pm$  0.003 &  0.442 &  0.14 &  0.61 &  1.219 \\
 1.988 & 10.0   &  2.714 $\pm$  0.014 & -0.039 $\pm$  0.003 &  0.239 $\pm$  0.003 &  0.432 &  0.26 &  0.54 &  0.776 \\
 1.988 & 11.3   &  3.063 $\pm$  0.014 &  0.062 $\pm$  0.003 &  0.179 $\pm$  0.003 &  0.460 &  0.26 &  0.55 &  0.908 \\
 1.988 & 12.8   &  3.068 $\pm$  0.014 &  0.070 $\pm$  0.002 &  0.172 $\pm$  0.002 &  0.458 &  0.27 &  0.54 &  0.888 \\
 1.988 & 15.0   &  3.079 $\pm$  0.014 &  0.072 $\pm$  0.002 &  0.172 $\pm$  0.002 &  0.454 &  0.28 &  0.53 &  0.855 \\
 1.988 & 18.0   &  3.172 $\pm$  0.013 &  0.029 $\pm$  0.002 &  0.226 $\pm$  0.002 &  0.438 &  0.35 &  0.49 &  0.724 \\
 1.988 & 21.0   &  3.368 $\pm$  0.013 &  0.017 $\pm$  0.002 &  0.261 $\pm$  0.002 &  0.424 &  0.40 &  0.45 &  0.651 \\
\hline 
\end{tabular}
\end{center}
\end{table*}
\begin{table*}
\begin{center}
\caption{Continuation of Table \ref{tab_jwst_2_all-a}}
\label{tab_jwst_2_all-c}
\begin{tabular}{ccccccccc}
\hline \hline
$\lambda_1$ (\um) & $\lambda_2$ (\um) & $c_0 \pm \Delta~c_0$ & $c_1 \pm \Delta~c_1$ & $c_2 \pm \Delta~c_2$ & $\sigma_{\delta}$ & $m_{\rm LvL}$ & $b_{\rm LvL}$ & $\sigma_{\delta}^{\rm updated}$ \\
\hline
11.3   & 21.0   &  4.300 $\pm$  0.015 & -0.076 $\pm$  0.004 &  0.455 $\pm$  0.004 &  0.525 &  0.57 &  0.33 &  0.655 \\
11.3   & 25.5   &  4.817 $\pm$  0.014 & -0.071 $\pm$  0.003 &  0.512 $\pm$  0.003 &  0.486 &  0.64 &  0.28 &  0.573 \\
12.8   & 15.0   &  4.142 $\pm$  0.016 &  0.248 $\pm$  0.013 &  0.101 $\pm$  0.013 &  0.571 &  0.47 &  0.39 &  0.770 \\
12.8   & 18.0   &  3.911 $\pm$  0.015 & -0.359 $\pm$  0.007 &  0.699 $\pm$  0.007 &  0.540 &  0.53 &  0.35 &  0.694 \\
12.8   & 21.0   &  4.185 $\pm$  0.014 & -0.234 $\pm$  0.005 &  0.608 $\pm$  0.004 &  0.518 &  0.57 &  0.32 &  0.646 \\
12.8   & 25.5   &  4.769 $\pm$  0.014 & -0.162 $\pm$  0.003 &  0.603 $\pm$  0.003 &  0.480 &  0.65 &  0.26 &  0.566 \\
15.0   & 18.0   &  3.761 $\pm$  0.014 & -1.311 $\pm$  0.011 &  1.651 $\pm$  0.011 &  0.504 &  0.57 &  0.32 &  0.640 \\
15.0   & 21.0   &  4.172 $\pm$  0.013 & -0.571 $\pm$  0.006 &  0.952 $\pm$  0.006 &  0.497 &  0.60 &  0.29 &  0.615 \\
15.0   & 25.5   &  4.826 $\pm$  0.013 & -0.310 $\pm$  0.003 &  0.763 $\pm$  0.004 &  0.465 &  0.67 &  0.24 &  0.544 \\
18.0   & 21.0   &  4.441 $\pm$  0.013 & -1.017 $\pm$  0.011 &  1.423 $\pm$  0.011 &  0.500 &  0.60 &  0.29 &  0.613 \\
18.0   & 25.5   &  5.043 $\pm$  0.013 & -0.431 $\pm$  0.004 &  0.906 $\pm$  0.005 &  0.463 &  0.68 &  0.23 &  0.537 \\
21.0   & 25.5   &  5.336 $\pm$  0.013 & -0.973 $\pm$  0.008 &  1.482 $\pm$  0.008 &  0.446 &  0.71 &  0.20 &  0.507 \\
\hline 
\end{tabular}
\end{center}
\end{table*}

\begin{table*}
\begin{center}
\caption{All possible combinations of three {\it JWST} filters}
\label{tab_jwst_3_all-a}
\begin{tabular}{ccccccccccc}
\hline \hline
$\lambda_1$ (\um) & $\lambda_2$ (\um) & $\lambda_3$ (\um) & $c_0 \pm \Delta~c_0$ & $c_1 \pm \Delta~c_1$ & $c_2 \pm \Delta~c_2$ & $c_3 \pm \Delta~c_3$ & $\sigma_{\delta}$ & $m_{\rm LvL}$ & $b_{\rm LvL}$ & $\sigma_{\delta}^{\rm updated}$\\
\hline
  0.704 &   0.903 &   1.154 &  -2.031 $\pm$ -12.502 &  -8.417 $\pm$   0.036 &  20.700 $\pm$   0.043 & -12.502 $\pm$   0.094 &   0.880 & -0.42 &  0.96 &   1.150 \\
  0.704 &   0.903 &   1.501 &  -0.625 $\pm$  -3.060 &  -3.789 $\pm$   0.025 &   6.760 $\pm$   0.028 &  -3.060 $\pm$   0.044 &   0.660 & -0.36 &  0.94 &   0.992 \\
  0.704 &   0.903 &   1.988 &   0.540 $\pm$  -0.424 &  -0.437 $\pm$   0.017 &   0.854 $\pm$   0.017 &  -0.424 $\pm$   0.024 &   0.449 & -0.07 &  0.74 &   1.829 \\
  0.704 &   0.903 &   2.776 &   1.702 $\pm$   0.805 &   2.541 $\pm$   0.013 &  -3.300 $\pm$   0.012 &   0.805 $\pm$   0.016 &   0.353 &  0.42 &  0.40 &   0.574 \\
  0.704 &   0.903 &   3.566 &   2.631 $\pm$   1.135 &   3.663 $\pm$   0.015 &  -4.698 $\pm$   0.013 &   1.135 $\pm$   0.017 &   0.415 &  0.70 &  0.22 &   0.553 \\
  0.704 &   0.903 &   4.401 &   3.546 $\pm$   1.411 &   4.549 $\pm$   0.020 &  -5.811 $\pm$   0.016 &   1.411 $\pm$   0.020 &   0.523 &  0.96 &  0.04 &   0.541 \\
  0.704 &   0.903 &     5.6 &   4.254 $\pm$   1.606 &   5.226 $\pm$   0.025 &  -6.659 $\pm$   0.019 &   1.606 $\pm$   0.023 &   0.650 &  1.25 & -0.14 &   0.521 \\
\hline 
\end{tabular}
\end{center}
\end{table*}

\begin{table*}
\begin{center}
\caption{Continuation of Table \ref{tab_jwst_3_all-a}}
\label{tab_jwst_3_all-b}
\begin{tabular}{ccccccccccc}
\hline \hline
$\lambda_1$ (\um) & $\lambda_2$ (\um) & $\lambda_3$ (\um) & $c_0 \pm \Delta~c_0$ & $c_1 \pm \Delta~c_1$ & $c_2 \pm \Delta~c_2$ & $c_3 \pm \Delta~c_3$ & $\sigma_{\delta}$ & $m_{\rm LvL}$ & $b_{\rm LvL}$ & $\sigma_{\delta}^{\rm updated}$\\
\hline
  0.704 &   2.776 &    18.0 &   1.911 $\pm$   0.602 &   0.173 $\pm$   0.011 &  -0.666 $\pm$   0.001 &   0.602 $\pm$   0.003 &   0.303 &  0.58 &  0.31 &   0.422 \\
  0.704 &   2.776 &    21.0 &   2.024 $\pm$   0.277 &   0.105 $\pm$   0.011 &  -0.247 $\pm$   0.001 &   0.277 $\pm$   0.003 &   0.317 &  0.33 &  0.48 &   0.425 \\
  0.704 &   2.776 &    25.5 &   1.721 $\pm$  -0.018 &   0.031 $\pm$   0.016 &   0.096 $\pm$   0.002 &  -0.018 $\pm$   0.003 &   0.427 &  0.05 &  0.68 &   2.201 \\
  0.704 &   3.566 &   4.401 &   8.555 $\pm$  27.785 &   1.407 $\pm$   0.060 & -28.789 $\pm$   0.007 &  27.785 $\pm$   0.097 &   1.132 &  1.68 & -0.31 &   0.707 \\
  0.704 &   3.566 &     5.6 &   7.272 $\pm$  17.999 &   1.562 $\pm$   0.067 & -19.456 $\pm$   0.008 &  17.999 $\pm$   0.065 &   1.407 &  2.48 & -0.79 &   0.614 \\
  0.704 &   3.566 &     7.7 &  -5.161 $\pm$  10.164 &   1.034 $\pm$   0.063 & -12.298 $\pm$   0.007 &  10.164 $\pm$   0.041 &   1.335 &  2.36 & -0.81 &   0.587 \\
  0.704 &   3.566 &    10.0 &   1.670 $\pm$   2.447 &  -0.068 $\pm$   0.022 &  -2.371 $\pm$   0.002 &   2.447 $\pm$   0.008 &   0.612 &  1.24 & -0.13 &   0.490 \\
\hline 
\end{tabular}
\end{center}
\end{table*}

\begin{table*}
\begin{center}
\caption{Continuation of Table \ref{tab_jwst_3_all-a}}
\label{tab_jwst_3_all-c}
\begin{tabular}{ccccccccccc}
\hline \hline
$\lambda_1$ (\um) & $\lambda_2$ (\um) & $\lambda_3$ (\um) & $c_0 \pm \Delta~c_0$ & $c_1 \pm \Delta~c_1$ & $c_2 \pm \Delta~c_2$ & $c_3 \pm \Delta~c_3$ & $\sigma_{\delta}$ & $m_{\rm LvL}$ & $b_{\rm LvL}$ & $\sigma_{\delta}^{\rm updated}$\\
\hline
  0.903 &   1.154 &   4.401 &   3.505 $\pm$   2.619 &   8.445 $\pm$   0.028 & -10.971 $\pm$   0.029 &   2.619 $\pm$   0.037 &   0.707 &  1.33 & -0.20 &   0.545 \\
  0.903 &   1.154 &     5.6 &   4.056 $\pm$   2.692 &   8.978 $\pm$   0.033 & -11.575 $\pm$   0.032 &   2.692 $\pm$   0.040 &   0.842 &  1.62 & -0.38 &   0.535 \\
  0.903 &   1.154 &     7.7 &   1.761 $\pm$   2.260 &   8.278 $\pm$   0.033 & -10.693 $\pm$   0.031 &   2.260 $\pm$   0.038 &   0.874 &  1.69 & -0.43 &   0.520 \\
  0.903 &   1.154 &    10.0 &   2.877 $\pm$   1.545 &   3.969 $\pm$   0.024 &  -5.430 $\pm$   0.017 &   1.545 $\pm$   0.021 &   0.652 &  1.43 & -0.26 &   0.448 \\
  0.903 &   1.154 &    11.3 &   2.590 $\pm$   1.305 &   3.429 $\pm$   0.022 &  -4.658 $\pm$   0.016 &   1.305 $\pm$   0.019 &   0.610 &  1.30 & -0.17 &   0.473 \\
  0.903 &   1.154 &    12.8 &   2.319 $\pm$   1.073 &   3.264 $\pm$   0.019 &  -4.270 $\pm$   0.014 &   1.073 $\pm$   0.017 &   0.539 &  1.17 & -0.09 &   0.465 \\
  0.903 &   1.154 &    15.0 &   2.057 $\pm$   0.846 &   2.898 $\pm$   0.016 &  -3.682 $\pm$   0.012 &   0.846 $\pm$   0.014 &   0.453 &  1.01 &  0.02 &   0.451 \\
\hline 
\end{tabular}
\end{center}
\end{table*}

\begin{table*}
\begin{center}
\caption{Continuation of Table \ref{tab_jwst_3_all-a}}
\label{tab_jwst_3_all-d}
\begin{tabular}{ccccccccccc}
\hline \hline
$\lambda_1$ (\um) & $\lambda_2$ (\um) & $\lambda_3$ (\um) & $c_0 \pm \Delta~c_0$ & $c_1 \pm \Delta~c_1$ & $c_2 \pm \Delta~c_2$ & $c_3 \pm \Delta~c_3$ & $\sigma_{\delta}$ & $m_{\rm LvL}$ & $b_{\rm LvL}$ & $\sigma_{\delta}^{\rm updated}$\\
\hline
  0.903 &   4.401 &    18.0 &   2.047 $\pm$   0.010 &   0.031 $\pm$   0.015 &   0.099 $\pm$   0.002 &   0.010 $\pm$   0.004 &   0.420 &  0.11 &  0.64 &   1.241 \\
  0.903 &   4.401 &    21.0 &   2.012 $\pm$  -0.246 &  -0.001 $\pm$   0.019 &   0.380 $\pm$   0.002 &  -0.246 $\pm$   0.004 &   0.526 & -0.08 &  0.77 &   2.162 \\
  0.903 &   4.401 &    25.5 &   1.251 $\pm$  -0.491 &  -0.045 $\pm$   0.025 &   0.578 $\pm$   0.003 &  -0.491 $\pm$   0.005 &   0.669 & -0.32 &  0.94 &   0.908 \\
  0.903 &     5.6 &     7.7 &  -7.678 $\pm$  10.321 &   0.216 $\pm$   0.051 & -11.530 $\pm$   0.004 &  10.321 $\pm$   0.042 &   0.856 &  0.86 &  0.16 &   0.935 \\
  0.903 &     5.6 &    10.0 &   1.974 $\pm$   0.612 &  -0.039 $\pm$   0.014 &  -0.449 $\pm$   0.002 &   0.612 $\pm$   0.005 &   0.372 &  0.34 &  0.48 &   0.550 \\
  0.903 &     5.6 &    11.3 &   2.041 $\pm$   0.413 &  -0.006 $\pm$   0.014 &  -0.273 $\pm$   0.002 &   0.413 $\pm$   0.005 &   0.391 &  0.29 &  0.52 &   0.608 \\
  0.903 &     5.6 &    12.8 &   2.089 $\pm$   0.234 &   0.031 $\pm$   0.014 &  -0.124 $\pm$   0.002 &   0.234 $\pm$   0.005 &   0.397 &  0.22 &  0.56 &   0.692 \\
\hline 
\end{tabular}
\end{center}
\end{table*}

\begin{table*}
\begin{center}
\caption{Continuation of Table \ref{tab_jwst_3_all-a}}
\label{tab_jwst_3_all-e}
\begin{tabular}{ccccccccccc}
\hline \hline
$\lambda_1$ (\um) & $\lambda_2$ (\um) & $\lambda_3$ (\um) & $c_0 \pm \Delta~c_0$ & $c_1 \pm \Delta~c_1$ & $c_2 \pm \Delta~c_2$ & $c_3 \pm \Delta~c_3$ & $\sigma_{\delta}$ & $m_{\rm LvL}$ & $b_{\rm LvL}$ & $\sigma_{\delta}^{\rm updated}$\\
\hline
  1.154 &   1.988 &    21.0 &   2.047 $\pm$   0.458 &   0.454 $\pm$   0.010 &  -0.781 $\pm$   0.003 &   0.458 $\pm$   0.004 &   0.287 &  0.51 &  0.35 &   0.398 \\
  1.154 &   1.988 &    25.5 &   2.014 $\pm$   0.169 &   0.146 $\pm$   0.013 &  -0.175 $\pm$   0.004 &   0.169 $\pm$   0.005 &   0.343 &  0.24 &  0.53 &   0.442 \\
  1.154 &   2.776 &   3.566 &  10.810 $\pm$  21.627 &   4.192 $\pm$   0.054 & -24.992 $\pm$   0.016 &  21.627 $\pm$   0.085 &   1.075 &  1.10 &  0.05 &   0.990 \\
  1.154 &   2.776 &   4.401 &  13.719 $\pm$  17.250 &   4.967 $\pm$   0.069 & -21.269 $\pm$   0.019 &  17.250 $\pm$   0.072 &   1.303 &  1.71 & -0.27 &   0.778 \\
  1.154 &   2.776 &     5.6 &  12.789 $\pm$  13.445 &   4.852 $\pm$   0.073 & -17.575 $\pm$   0.019 &  13.445 $\pm$   0.060 &   1.485 &  2.32 & -0.62 &   0.656 \\
  1.154 &   2.776 &     7.7 &   0.914 $\pm$   8.617 &   3.562 $\pm$   0.059 & -12.645 $\pm$   0.015 &   8.617 $\pm$   0.043 &   1.405 &  2.34 & -0.72 &   0.627 \\
  1.154 &   2.776 &    10.0 &   2.956 $\pm$   2.609 &   0.386 $\pm$   0.026 &  -2.889 $\pm$   0.005 &   2.609 $\pm$   0.010 &   0.743 &  1.44 & -0.23 &   0.503 \\
\hline 
\end{tabular}
\end{center}
\end{table*}

\begin{table*}
\begin{center}
\caption{Continuation of Table \ref{tab_jwst_3_all-a}}
\label{tab_jwst_3_all-f}
\begin{tabular}{ccccccccccc}
\hline \hline
$\lambda_1$ (\um) & $\lambda_2$ (\um) & $\lambda_3$ (\um) & $c_0 \pm \Delta~c_0$ & $c_1 \pm \Delta~c_1$ & $c_2 \pm \Delta~c_2$ & $c_3 \pm \Delta~c_3$ & $\sigma_{\delta}$ & $m_{\rm LvL}$ & $b_{\rm LvL}$ & $\sigma_{\delta}^{\rm updated}$\\
\hline
  1.154 &    15.0 &    18.0 &   2.788 $\pm$  -4.217 &  -0.024 $\pm$   0.031 &   4.379 $\pm$   0.004 &  -4.217 $\pm$   0.021 &   0.939 & -0.10 &  0.83 &   4.583 \\
  1.154 &    15.0 &    21.0 &   1.397 $\pm$  -2.261 &  -0.128 $\pm$   0.031 &   2.385 $\pm$   0.004 &  -2.261 $\pm$   0.012 &   0.927 & -0.12 &  0.86 &   5.203 \\
  1.154 &    15.0 &    25.5 &  -0.400 $\pm$  -1.687 &  -0.177 $\pm$   0.036 &   1.668 $\pm$   0.004 &  -1.687 $\pm$   0.008 &   0.982 & -0.27 &  0.97 &   2.548 \\
  1.154 &    18.0 &    21.0 &   0.639 $\pm$  -4.178 &  -0.291 $\pm$   0.029 &   4.397 $\pm$   0.004 &  -4.178 $\pm$   0.022 &   0.844 &  0.07 &  0.75 &   5.286 \\
  1.154 &    18.0 &    25.5 &  -0.914 $\pm$  -2.170 &  -0.293 $\pm$   0.035 &   2.224 $\pm$   0.004 &  -2.170 $\pm$   0.011 &   0.897 & -0.10 &  0.87 &   6.156 \\
  1.154 &    21.0 &    25.5 &  -1.236 $\pm$  -3.716 &  -0.313 $\pm$   0.036 &   3.759 $\pm$   0.004 &  -3.716 $\pm$   0.019 &   0.866 & -0.05 &  0.83 &   5.190 \\
  1.501 &   1.988 &   2.776 &   0.138 $\pm$  13.387 &  11.714 $\pm$   0.040 & -25.333 $\pm$   0.046 &  13.387 $\pm$   0.094 &   0.746 &  0.94 &  0.19 &   0.775 \\
\hline 
\end{tabular}
\end{center}
\end{table*}

\begin{table*}
\begin{center}
\caption{Continuation of Table \ref{tab_jwst_3_all-a}}
\label{tab_jwst_3_all-g}
\begin{tabular}{ccccccccccc}
\hline \hline
$\lambda_1$ (\um) & $\lambda_2$ (\um) & $\lambda_3$ (\um) & $c_0 \pm \Delta~c_0$ & $c_1 \pm \Delta~c_1$ & $c_2 \pm \Delta~c_2$ & $c_3 \pm \Delta~c_3$ & $\sigma_{\delta}$ & $m_{\rm LvL}$ & $b_{\rm LvL}$ & $\sigma_{\delta}^{\rm updated}$\\
\hline
  1.501 &     7.7 &    25.5 &  -0.039 $\pm$  -1.124 &  -0.066 $\pm$   0.036 &   1.048 $\pm$   0.005 &  -1.124 $\pm$   0.007 &   1.004 & -0.63 &  1.17 &   0.932 \\
  1.501 &    10.0 &    11.3 &   4.029 $\pm$  -3.078 &  -0.079 $\pm$   0.020 &   3.483 $\pm$   0.003 &  -3.078 $\pm$   0.034 &   0.496 &  0.25 &  0.56 &   1.107 \\
  1.501 &    10.0 &    12.8 &   3.539 $\pm$  -1.278 &  -0.165 $\pm$   0.018 &   1.713 $\pm$   0.003 &  -1.278 $\pm$   0.016 &   0.508 &  0.27 &  0.56 &   1.082 \\
  1.501 &    10.0 &    15.0 &   3.398 $\pm$  -0.791 &  -0.167 $\pm$   0.018 &   1.213 $\pm$   0.003 &  -0.791 $\pm$   0.010 &   0.533 &  0.26 &  0.56 &   1.218 \\
  1.501 &    10.0 &    18.0 &   3.509 $\pm$  -1.123 &  -0.210 $\pm$   0.024 &   1.575 $\pm$   0.004 &  -1.123 $\pm$   0.009 &   0.715 &  0.13 &  0.66 &   3.785 \\
  1.501 &    10.0 &    21.0 &   2.779 $\pm$  -1.004 &  -0.200 $\pm$   0.026 &   1.364 $\pm$   0.004 &  -1.004 $\pm$   0.008 &   0.786 &  0.03 &  0.73 &   6.364 \\
  1.501 &    10.0 &    25.5 &   1.462 $\pm$  -1.000 &  -0.196 $\pm$   0.031 &   1.210 $\pm$   0.005 &  -1.000 $\pm$   0.008 &   0.885 & -0.13 &  0.85 &   4.382 \\
\hline 
\end{tabular}
\end{center}
\end{table*}

\begin{table*}
\begin{center}
\caption{Continuation of Table \ref{tab_jwst_3_all-a}}
\label{tab_jwst_3_all-h}
\begin{tabular}{ccccccccccc}
\hline \hline
$\lambda_1$ (\um) & $\lambda_2$ (\um) & $\lambda_3$ (\um) & $c_0 \pm \Delta~c_0$ & $c_1 \pm \Delta~c_1$ & $c_2 \pm \Delta~c_2$ & $c_3 \pm \Delta~c_3$ & $\sigma_{\delta}$ & $m_{\rm LvL}$ & $b_{\rm LvL}$ & $\sigma_{\delta}^{\rm updated}$\\
\hline
  1.988 &     7.7 &    12.8 &   2.082 $\pm$  -0.768 &   0.061 $\pm$   0.019 &   0.842 $\pm$   0.004 &  -0.768 $\pm$   0.008 &   0.584 & -0.10 &  0.76 &   2.543 \\
  1.988 &     7.7 &    15.0 &   2.164 $\pm$  -0.771 &  -0.036 $\pm$   0.021 &   0.940 $\pm$   0.004 &  -0.771 $\pm$   0.008 &   0.625 & -0.12 &  0.78 &   2.387 \\
  1.988 &     7.7 &    18.0 &   2.067 $\pm$  -0.995 &  -0.011 $\pm$   0.026 &   1.111 $\pm$   0.005 &  -0.995 $\pm$   0.008 &   0.786 & -0.31 &  0.92 &   1.447 \\
  1.988 &     7.7 &    21.0 &   1.408 $\pm$  -1.029 &  -0.056 $\pm$   0.029 &   1.108 $\pm$   0.006 &  -1.029 $\pm$   0.008 &   0.877 & -0.42 &  1.01 &   1.218 \\
  1.988 &     7.7 &    25.5 &  -0.109 $\pm$  -1.111 &  -0.103 $\pm$   0.035 &   1.063 $\pm$   0.006 &  -1.111 $\pm$   0.008 &   0.993 & -0.61 &  1.16 &   0.952 \\
  1.988 &    10.0 &    11.3 &   3.882 $\pm$  -3.127 &  -0.164 $\pm$   0.020 &   3.598 $\pm$   0.003 &  -3.127 $\pm$   0.033 &   0.484 &  0.26 &  0.57 &   1.067 \\
  1.988 &    10.0 &    12.8 &   3.332 $\pm$  -1.248 &  -0.252 $\pm$   0.018 &   1.744 $\pm$   0.004 &  -1.248 $\pm$   0.014 &   0.492 &  0.27 &  0.57 &   1.059 \\
\hline 
\end{tabular}
\end{center}
\end{table*}

\begin{table*}
\begin{center}
\caption{Continuation of Table \ref{tab_jwst_3_all-a}}
\label{tab_jwst_3_all-i}
\begin{tabular}{ccccccccccc}
\hline \hline
$\lambda_1$ (\um) & $\lambda_2$ (\um) & $\lambda_3$ (\um) & $c_0 \pm \Delta~c_0$ & $c_1 \pm \Delta~c_1$ & $c_2 \pm \Delta~c_2$ & $c_3 \pm \Delta~c_3$ & $\sigma_{\delta}$ & $m_{\rm LvL}$ & $b_{\rm LvL}$ & $\sigma_{\delta}^{\rm updated}$\\
\hline
  2.776 &    10.0 &    15.0 &   3.242 $\pm$  -0.682 &  -0.254 $\pm$   0.018 &   1.174 $\pm$   0.004 &  -0.682 $\pm$   0.009 &   0.514 &  0.24 &  0.60 &   1.263 \\
  2.776 &    10.0 &    18.0 &   3.431 $\pm$  -1.053 &  -0.293 $\pm$   0.024 &   1.581 $\pm$   0.006 &  -1.053 $\pm$   0.010 &   0.699 &  0.08 &  0.74 &   4.471 \\
  2.776 &    10.0 &    21.0 &   2.725 $\pm$  -0.955 &  -0.272 $\pm$   0.026 &   1.384 $\pm$   0.007 &  -0.955 $\pm$   0.009 &   0.770 & -0.03 &  0.82 &   6.515 \\
  2.776 &    10.0 &    25.5 &   1.398 $\pm$  -0.972 &  -0.272 $\pm$   0.030 &   1.253 $\pm$   0.007 &  -0.972 $\pm$   0.009 &   0.872 & -0.21 &  0.95 &   2.810 \\
  2.776 &    11.3 &    12.8 &   3.370 $\pm$  -1.270 &  -0.073 $\pm$   0.016 &   1.605 $\pm$   0.004 &  -1.270 $\pm$   0.017 &   0.512 &  0.26 &  0.58 &   1.141 \\
  2.776 &    11.3 &    15.0 &   3.423 $\pm$  -0.751 &  -0.080 $\pm$   0.017 &   1.096 $\pm$   0.004 &  -0.751 $\pm$   0.010 &   0.541 &  0.24 &  0.60 &   1.362 \\
  2.776 &    11.3 &    18.0 &   3.571 $\pm$  -1.158 &  -0.107 $\pm$   0.023 &   1.522 $\pm$   0.005 &  -1.158 $\pm$   0.010 &   0.731 &  0.07 &  0.74 &   4.848 \\
\hline 
\end{tabular}
\end{center}
\end{table*}

\begin{table*}
\begin{center}
\caption{Continuation of Table \ref{tab_jwst_3_all-a}}
\label{tab_jwst_3_all-j}
\begin{tabular}{ccccccccccc}
\hline \hline
$\lambda_1$ (\um) & $\lambda_2$ (\um) & $\lambda_3$ (\um) & $c_0 \pm \Delta~c_0$ & $c_1 \pm \Delta~c_1$ & $c_2 \pm \Delta~c_2$ & $c_3 \pm \Delta~c_3$ & $\sigma_{\delta}$ & $m_{\rm LvL}$ & $b_{\rm LvL}$ & $\sigma_{\delta}^{\rm updated}$\\
\hline
  3.566 &    18.0 &    25.5 &  -0.367 $\pm$  -1.974 &  -0.518 $\pm$   0.033 &   2.320 $\pm$   0.006 &  -1.974 $\pm$   0.012 &   0.873 &  0.09 &  0.73 &   6.647 \\
  3.566 &    21.0 &    25.5 &  -1.049 $\pm$  -3.695 &  -0.642 $\pm$   0.035 &   4.097 $\pm$   0.006 &  -3.695 $\pm$   0.021 &   0.865 &  0.18 &  0.65 &   3.623 \\
  4.401 &     5.6 &     7.7 &  -7.111 $\pm$  10.615 &   9.440 $\pm$   0.053 & -20.856 $\pm$   0.046 &  10.615 $\pm$   0.083 &   0.891 &  0.36 &  0.46 &   2.150 \\
  4.401 &     5.6 &    10.0 &   2.197 $\pm$   0.566 &   0.919 $\pm$   0.014 &  -1.322 $\pm$   0.012 &   0.566 $\pm$   0.013 &   0.405 &  0.24 &  0.55 &   0.898 \\
  4.401 &     5.6 &    11.3 &   2.181 $\pm$   0.437 &   0.934 $\pm$   0.014 &  -1.209 $\pm$   0.012 &   0.437 $\pm$   0.013 &   0.415 &  0.23 &  0.55 &   0.920 \\
  4.401 &     5.6 &    12.8 &   2.090 $\pm$   0.342 &   1.073 $\pm$   0.014 &  -1.257 $\pm$   0.012 &   0.342 $\pm$   0.014 &   0.411 &  0.21 &  0.56 &   0.950 \\
  4.401 &     5.6 &    15.0 &   2.215 $\pm$   0.095 &   0.865 $\pm$   0.016 &  -0.791 $\pm$   0.013 &   0.095 $\pm$   0.016 &   0.436 &  0.12 &  0.61 &   1.527 \\
\hline 
\end{tabular}
\end{center}
\end{table*}

\begin{table*}
\begin{center}
\caption{Continuation of Table \ref{tab_jwst_3_all-a}}
\label{tab_jwst_3_all-k}
\begin{tabular}{ccccccccccc}
\hline \hline
$\lambda_1$ (\um) & $\lambda_2$ (\um) & $\lambda_3$ (\um) & $c_0 \pm \Delta~c_0$ & $c_1 \pm \Delta~c_1$ & $c_2 \pm \Delta~c_2$ & $c_3 \pm \Delta~c_3$ & $\sigma_{\delta}$ & $m_{\rm LvL}$ & $b_{\rm LvL}$ & $\sigma_{\delta}^{\rm updated}$\\
\hline
    5.6 &    15.0 &    25.5 &   0.430 $\pm$  -1.449 &  -0.326 $\pm$   0.034 &   1.686 $\pm$   0.008 &  -1.449 $\pm$   0.011 &   0.956 & -0.09 &  0.83 &   6.208 \\
    5.6 &    18.0 &    21.0 &   0.877 $\pm$  -3.952 &  -0.628 $\pm$   0.029 &   4.560 $\pm$   0.007 &  -3.952 $\pm$   0.024 &   0.845 &  0.20 &  0.66 &   3.146 \\
    5.6 &    18.0 &    25.5 &  -0.215 $\pm$  -1.894 &  -0.558 $\pm$   0.033 &   2.316 $\pm$   0.007 &  -1.894 $\pm$   0.013 &   0.872 &  0.07 &  0.75 &   5.920 \\
    5.6 &    21.0 &    25.5 &  -0.900 $\pm$  -3.579 &  -0.696 $\pm$   0.036 &   4.075 $\pm$   0.006 &  -3.579 $\pm$   0.021 &   0.868 &  0.12 &  0.72 &   5.172 \\
    7.7 &    10.0 &    11.3 &   4.131 $\pm$  -2.701 &  -0.390 $\pm$   0.019 &   3.436 $\pm$   0.006 &  -2.701 $\pm$   0.029 &   0.484 &  0.28 &  0.56 &   1.063 \\
    7.7 &    10.0 &    12.8 &   3.449 $\pm$  -0.693 &  -0.268 $\pm$   0.016 &   1.237 $\pm$   0.006 &  -0.693 $\pm$   0.010 &   0.479 &  0.26 &  0.57 &   1.053 \\
    7.7 &    10.0 &    15.0 &   3.393 $\pm$  -0.474 &  -0.236 $\pm$   0.017 &   0.979 $\pm$   0.006 &  -0.474 $\pm$   0.007 &   0.499 &  0.24 &  0.58 &   1.217 \\
\hline 
\end{tabular}
\end{center}
\end{table*}


\bsp	
\label{lastpage}
\end{document}